\documentclass[showpacs,aps,prd,preprint,nofootinbib,showkeys,unsortedaddress,raggedbottom]{revtex4-1}
\pdfoutput=1

\usepackage{bm}
\usepackage{amsmath}
\usepackage{graphicx}
\usepackage{subfigure}
\usepackage[usenames,dvipsnames]{color}
\definecolor{darkblue}{RGB}{0,0,196}
\usepackage[colorlinks=true,linkcolor=darkblue,citecolor=darkblue,urlcolor=darkblue]{hyperref}

\def\be{\begin{equation}}
\def\ee{\end{equation}}

\def\ba{\begin{eqnarray}}
\def\ea{\end{eqnarray}}

\begin{document}

\title{Leading-order anisotropic hydrodynamics for central collisions}

\author{Mohammad Nopoush and Michael Strickland}
\affiliation{Kent State University, Kent, OH 44242 USA}

\author{Radoslaw Ryblewski}
\affiliation {The H. Niewodnicza\'nski Institute of Nuclear Physics, Polish Academy of Sciences, PL-31342 Krak\'ow, Poland}

\author{Dennis Bazow, Ulrich Heinz, and Mauricio Martinez}
\affiliation{Department of Physics, The Ohio State University, Columbus, Ohio 43210-1117, USA}

\begin{abstract}
We use leading-order anisotropic hydrodynamics to study an azimuthally-symmetric boost-invariant quark-gluon plasma.  We impose a realistic lattice-based equation of state and perform self-consistent anisotropic freeze-out to hadronic degrees of freedom.  We then compare our results for the full spatiotemporal evolution of the quark-gluon plasma and its subsequent freeze-out to results obtained using 1+1d Israel-Stewart second-order viscous hydrodynamics.  We find that for small shear viscosities, $4 \pi \eta/s \sim 1$, the two methods agree well for nucleus-nucleus collisions, however, for large shear viscosity to entropy density ratios or proton-nucleus collisions we find important corrections to the Israel-Stewart results for the final particle spectra and the total number of charged particles.  Finally, we demonstrate that the total number of charged particles produced is a monotonically increasing function of $4 \pi \eta/s$ in Israel-Stewart viscous hydrodynamics whereas in anisotropic hydrodynamics it has a maximum at $4 \pi \eta/s \sim 10$.  For all $4 \pi \eta/s > 0$, we find that for Pb-Pb collisions Israel-Stewart viscous hydrodynamics predicts more dissipative particle production than anisotropic hydrodynamics.
\end{abstract}

\maketitle

\section{Introduction}
\label{sec:introduction}

In recent years there have been significant advances in our understanding of the theory of relativistic hydrodynamics and its application to describing the spacetime evolution of the quark-gluon plasma created in ultrarelativistic heavy-ion collisions.  Early works applied relativistic ideal hydrodynamics \cite{Huovinen:2001cy,Hirano:2002ds,Kolb:2003dz} and later works developed and applied second-order viscous hydrodynamics \cite{Muronga:2001zk,Muronga:2003ta,Muronga:2004sf,Heinz:2005bw,Baier:2006um,Romatschke:2007mq,Baier:2007ix,Dusling:2007gi,Luzum:2008cw,Song:2008hj,Heinz:2009xj,Bozek:2009ty,Bozek:2009dw,El:2009vj,PeraltaRamos:2009kg,PeraltaRamos:2010je,Denicol:2010tr,Denicol:2010xn,Schenke:2010rr,Schenke:2011tv,Bozek:2011wa,Bozek:2011ua,Niemi:2011ix,Niemi:2012ry,Bozek:2012qs,Denicol:2012cn,Denicol:2012es,PeraltaRamos:2012xk,Calzetta:2014hra,Denicol:2014vaa,Florkowski:2015lra,Ryu:2015vwa}.  In recent years, the framework of anisotropic hydrodynamics was developed in order to better account for the large momentum-space anisotropies generated in relativistic heavy-ion collisions \cite{Martinez:2010sc,Florkowski:2010cf,Ryblewski:2010bs,Martinez:2010sd,Ryblewski:2011aq,Florkowski:2011jg,Martinez:2012tu,Ryblewski:2012rr,Florkowski:2012as,Florkowski:2013uqa,Ryblewski:2013jsa,Bazow:2013ifa,Tinti:2013vba,Florkowski:2014bba,Florkowski:2014txa,Nopoush:2014pfa,Denicol:2014mca,Nopoush:2014pfa,Bazow:2015cha}.  This framework has been shown to more accurately describe the evolution of systems subject to boost-invariant and transversely homogeneous 0+1d flow than traditional viscous hydrodynamics approaches \cite{Florkowski:2013lza,Florkowski:2013lya,Bazow:2013ifa,Florkowski:2014bba,Nopoush:2014pfa,Florkowski:2014sfa,Bazow:2015cha} and has recently been shown to best reproduce exact solutions of Boltzmann equation subject to 1+1d Gubser flow \cite{Denicol:2014xca,Denicol:2014tha,Nopoush:2014qba}.  In its latest form, leading-order anisotropic hydrodynamics allows for multiple local momentum-space anisotropies in the argument of the non-equilibrium distribution function \cite{Martinez:2012tu,Tinti:2013vba,Nopoush:2014qba}.  In addition, it has been shown that it is possible to account for non-spheroidal/-ellipsoidal corrections via a modified shear correction resulting in ``viscous anisotropic hydrodynamics'' \cite{Bazow:2013ifa,Bazow:2015cha}.  With these improvements, anisotropic hydrodyanmics has been shown to work extremely well when compared to exact solutions of the massless and massive Boltzmann equation in relaxation time approximation \cite{Florkowski:2013lya,Bazow:2015cha,Florkowski:2013lza,Florkowski:2014sfa,Bazow:2013ifa}.

Despite this promise, turning anisotropic hydrodynamics into a practical phenomenological tool for use in modeling heavy-ion collisions requires two additional fundamental ingredients to be implemented:  (1) a realistic lattice-based equation of state (EoS) and (2) self-consistent anisotropic freeze-out to hadronic degrees of freedom.  For the EoS, it is not obvious a priori how one enforces thermodynamic relations in an anisotropic system.  As we will show, it is possible to impose the EoS as a relation between the isotropic energy density and pressure.  For anisotropic freeze-out, we determine the freeze-out hypersurface by specifying a critical energy density and then we use the leading-order anisotropic distribution function to compute particle spectra using ``anisotropic Cooper-Frye freeze-out''.  This method includes the leading dissipative corrections at freeze-out in a way that guarantees a positive definite one-particle distribution function at all momenta, thus avoiding the problem of regions of phase space where the distribution function is negative. We here restrict our anisotropic hydrodynamic analysis to the inclusion at leading order of the most important local momentum anisotropies \cite{Martinez:2012tu,Tinti:2013vba,Nopoush:2014qba}. The generalization to ``viscous anisotropic hydrodynamicsÕÕ \cite{Bazow:2013ifa,Bazow:2015cha} which also includes the smaller residual deviations from the leading-order distribution will be pursued in a future work.

As preparation for implementing the necessary ingredients in a 3+1d anisotropic hydrodynamics code, in this paper we perform the somewhat simpler task of implementing 1+1d anisotropic hydrodynamics for an azimuthally-symmetric and boost-invariant system.  We will compare to predictions of a 1+1d Israel-Stewart viscous hydrodynamics code in which we have implemented exactly the same initial conditions, freeze-out hypersurface algorithm, etc.  Since the 1+1d task is more straightforward computationally, this allows us to virtually eliminate systematic computational errors.  We compare our results for the full spatiotemporal evolution of the quark-gluon plasma and its subsequent freeze-out to results obtained with 1+1d Israel-Stewart second-order viscous hydrodynamics.  We find that for small shear viscosities, $4 \pi \eta/s \sim 1$, the two methods agree well for nucleus-nucleus collisions, however, for large shear viscosity to entropy density ratios or proton-nucleus collisions we find important corrections to the Israel-Stewart results for the final particle spectra and the total number of charged particles.  We demonstrate that the total number of charged particles is a monotonically increasing function of $4 \pi \eta/s$ in Israel-Stewart viscous hydrodynamics whereas in anisotropic hydrodynamics it peaks at $4 \pi \eta/s \sim 10$.  For all $4 \pi \eta/s > 0$ we find that for Pb-Pb collisions Israel-Stewart viscous hydrodynamics predicts harder spectra and more dissipative particle production than anisotropic hydrodynamics.

The structure of paper is as follows.  In Sec.~\ref{sec:conventions}, we specify the conventions used in the body of the text.  In Secs.~\ref{sec:setup} and \ref{sec:equations}, we present the setup and details for the conformal anisotropic hydrodynamics dynamical equations.  In Sec.~\ref{sec:eos}, we discuss how we implement a realistic EoS in the context of anisotropic hydrodynamics.  In Sec.~\ref{sec:freeze-out} we discuss how to implement anisotropic Cooper-Frye freeze-out in the context of leading-order anisotropic hydrodynamics.  In Sec.~\ref{sec:results} we present our numerical results for Pb-Pb and p-Pb collisions at LHC energies.  In Sec.~\ref{sec:conclusion} we summarize our findings, state our conclusions, and present an outlook for the future.  Finally, in six appendices we present details behind many of the derivations and results presented in the body of the text.

\section{Conventions}
\label{sec:conventions}

In this paper, the metric is taken to be ``mostly minus'' such that, in Minkowski space with $x^\mu=(t,x,y,z)$, the line element is
\be
ds^2=g_{\mu\nu}dx^\mu dx^\nu=dt^2-dx^2-dy^2-dz^2\,.
\ee
Since we deal with a system that is boost invariant along the beam-line we transform to a new variables defined by $\tau =\sqrt{t^2 - z^2}$ as the longitudinal proper time, and $\varsigma ={\rm tanh}^{-1}(z/t)$ as the longitudinal spacetime rapidity. Also, since the system is cylindrically symmetric with respect to the beam-line it is convenient to transform to the polar coordinates in the transverse plane with $r=\sqrt{x^{2}{+}y^{2}}$ and $\phi ={\rm tan}^{-1}(y/x)$. The new set of coordinates $x^\mu=(\tau,r,\phi,\varsigma)$ defines the polar Milne coordinates.

We also mention that the notations $A^{(\mu\nu)}\equiv\frac{1}{2}\left(A^{\mu\nu}{+}A^{\nu\mu}\right)$ and $A^{[\mu\nu]}\equiv\frac{1}{2}\left(A^{\mu\nu}{-}A^{\nu\mu}\right)$ denote symmetrization and antisymmetrization, respectively, and $A^{\langle \mu \nu\rangle}\equiv\Delta^{\mu\nu}_{\alpha\beta}A^{\alpha\beta}$ where $\Delta^{\mu\nu}_{\alpha\beta}\equiv\Delta^{(\mu}_\alpha\Delta^{\nu)}_\beta-\Delta^{\mu\nu}\Delta_{\alpha\beta}/3$ with $\Delta^{\mu\nu} \equiv g^{\mu\nu} - u^\mu u^\nu$.  The four-index projector $\Delta^{\mu\nu}_{\alpha\beta}$ projects out components of a rank-2 tensor which are traceless and transverse to the flow velocity $u$.

\section{Anisotropic hydrodynamics setup and basis vectors}
\label{sec:setup}

For the purposes of this paper, we assume that the system is boost invariant and azimuthally symmetric so that we can apply 1+1d hydrodynamical evolution.  In the case of anisotropic hydrodynamics we use an ellipsoidal form for the local-rest-frame (LRF) one-particle distribution function.  This form can be obtained by introducing an anisotropy tensor of the form \cite{Martinez:2012tu,Tinti:2013vba,Nopoush:2014qba}
\be 
\Xi^{\mu\nu}=u^\mu u^\nu+\xi^{\mu\nu} -  \Phi \Delta^{\mu\nu} \, ,
\label{eq:aniso-tensor1}
\ee
where $u^\mu$ is the four-velocity, $\xi^{\mu\nu}$ is a symmetric traceless anisotropy tensor, $\Phi$ is a parameter associated with the bulk viscous correction.  In the LRF, the anisotropy tensor $\xi^{\mu\nu}$ is $\xi^{\mu\nu}_{\rm LRF}\equiv \text{diag}(0,\xi_x,\xi_y,\xi_z)$ with $\xi^\mu_{\mu,\rm LRF} = \xi_x + \xi_y + \xi_z = 0$ \cite{Tinti:2013vba,Nopoush:2014qba}.\footnote{The labels $x$, $y$, and $z$ do not imply the cartesian components, but are merely labels for spacelike directions.  In this work, the vectors corresponding to $x$, $y$, and $z$ will be mapped to the radial, azimuthal, and rapidity directions, respectively.}  As mentioned above, the field $\Phi$ accounts for bulk viscous effects.  In the case of a massless (conformal) system, one can take $\Phi=0$.  We will assume from here on that, even in the non-conformal case, one has $\Phi=0$.  This assumption is analogous to assuming that the bulk correction to the pressure can be neglected relative to the shear correction in second-order viscous hydrodynamics.  Bulk viscous effects will be included in a future work. 

In general, $\xi^{\mu\nu}$ can have off-diagonal elements; however, based on the proper-time evolution of the shear-stress tensor components in simulations, one can expect that the off-diagonal elements are much smaller than the diagonal ones and can therefore be considered perturbatively~\cite{Song:2009gc}.  Importantly, however, for 1+1d expansion, similar to viscous hydrodynamics where it suffices to include only the diagonal contributions to the shear tensor ($\pi^r_r$, $\pi^\phi_\phi$, and $\pi^\varsigma_\varsigma$), in anisotropic hydrodynamics one only needs the diagonal components of the anisotropy tensor to describe a 1+1d system.  For a discussion of 1+1d second-order viscous hydrodynamics, see App.~\ref{app:vhydro}.

In a general frame one can expand the (diagonal) anisotropy tensor in covariant form
\be
\xi^{\mu\nu} = \xi_x X^\mu X^\nu+\xi_y Y^\mu Y^\nu+\xi_z Z^\mu Z^\nu \, ,
\label{eq:aniso-tensor2}
\ee
with
\be
\xi^{\mu}_{\ \mu} = 0 \, .
\label{eq:tracelessness}
\ee
The orthogonal basis vectors $u^\mu$, $X^\mu$, $Y^\mu$, and $Z^\mu$ reduce in the LRF to unit vectors in the $t$, $r$, $\phi$, and $z$ directions.  The basis vectors are explicitly defined in App.~\ref{app:basis}.  They obey the normalization conditions $u^\mu u_\mu =1$ and $X^\mu X_\mu = Y^\mu Y_\mu = Z^\mu Z_\mu = -1$, therefore, ${\Xi^\mu}_{\mu} = 1$.   In lab frame, $u^\mu$ and $X^\mu$ are unit vectors which point in a mixture of the $\tau$ and $r$ directions, $Y^\mu$ is a unit vector pointing in the azimuthal direction, and $Z^\mu$ is a unit vector pointing in the spatial rapidity direction. 

Using the tensor $\Xi^{\mu\nu}$, one can construct an anisotropic distribution function
\be
f(x,p)=f_{\rm eq}\!\left(\frac{1}{\lambda}\sqrt{p_\mu\Xi^{\mu\nu} p_\nu}\right) ,
\label{eq:pdf}
\ee
where $f_{\rm eq}(x) = 1/[\exp(x) + a]$ with $a = -1, +1,$ or 0 for Bose, Fermi, and Boltzmann statistics, respectively.  Above, $\lambda$ is a temperature-like scale that can be identified with the temperature, $T$, only when $\xi^{\mu\nu}=0$.\footnote{We assume herein that the chemical potential is zero.}

\section{1+1d anisotropic hydrodynamics}
\label{sec:equations}

We will obtain the anisotropic hydrodynamics equations by taking moments of Boltzmann equation in the relaxation time approximation (RTA). The Boltzmann equation in RTA is	
\be 
p^\mu d_\mu f = \frac{p^\mu u_\mu}{\tau_{\rm eq}}(f-f_{\rm eq})\, ,
\label{eq:boltzmanneq}
\ee
where $d_\mu$ is the covariant derivative which becomes the ordinary derivative $\partial_\mu$ in flat spacetimes and $\tau_{\rm eq}$ is the relaxation time. The right-hand side of Boltzmann equation is the collisional kernel which contains all interactions involved in the dynamics. If the particles comprising the fluid are massless, conformal invariance requires that $\tau_{\rm eq}$ is inversely proportional to the temperature, i.e. $\tau_{\rm eq} \propto 1/T$. In conformal RTA one has $\tau_{\rm eq} = 5\bar\eta/T$, where $\bar\eta=\eta/s$ with $\eta$ and $s$ being shear viscosity and entropy density, respectively. 

The general equations governing conformal 1+1d anisotropic hydrodynamics for a azimuthally-symmetric and boost-invariant system were obtained originally by Tinti et al \cite{Tinti:2013vba}.  The resulting dynamical equations give the evolution of six fields $\xi_x$, $\xi_y$, $\xi_z$, $\lambda$, $T$, and $\theta_\perp$ where $\theta_\perp = \tanh^{-1}(u^r/u^\tau)$ is the transverse rapidity.  Using the constraint (\ref{eq:tracelessness}) one is left with five independent parameters.  Taking four equations from the first and second moments of Boltzmann equation one has \cite{Tinti:2013vba,Nopoush:2014qba}
\ba
D_u\varepsilon +\varepsilon\theta_u+ P_x D_x\theta_\perp+P_y\frac{\sinh\theta_\perp}{r}+P_z\frac{\cosh \theta_\perp}{\tau}&=&0 \, ,
  \label{eq:equation-1}
\\
D_x P_x+P_x \theta_x+\varepsilon  D_u\theta_\perp-P_y \frac{\cosh \theta_\perp}{r}-P_z\frac{
   \sinh \theta_\perp}{\tau}&=&0 \,  , 
\label{eq:equation-2}
\\
\frac{D_u\alpha_i}{\alpha_i}-\frac{1}{3}\sum\limits_{j=x,y,z}\frac{D_u\alpha_j}{\alpha_j}-\sigma_i+\frac{1}{2\tau_{\rm eq}}\left(1-\frac{1}{\alpha_i^2}\right)\left(\frac{T}{\lambda}\right)^5\frac{1}{\alpha_x\alpha_y\alpha_z}&=&0 \, , \qquad i\in \{x,y\} \, .
\label{eq:equation-3}
\ea
Above $\alpha_i=1/\sqrt{1+\xi_i}\,$, $\varepsilon$ is the energy density, and $P_i$ are the spacelike diagonal components of the energy momentum tensor (pressures) with \cite{Nopoush:2014qba}\,\footnote{For non-conformal systems, one has instead $\alpha_i = 1/\sqrt{1+\xi_i+\Phi}$.}
\ba
\varepsilon &=& \varepsilon_{\rm eq}(\lambda){\cal R}(\alpha_x,\alpha_z) \, , \label{eq:ed} \\
P_x &=& P_{\rm eq} (\lambda){\cal H}_{Tx}(\alpha_x,\alpha_z) \, , \label{eq:px} \\
P_y &=& P_{\rm eq}(\lambda) {\cal H}_{Ty}(\alpha_x,\alpha_z) \, , \label{eq:py} \\
P_z &=& P_{\rm eq}(\lambda) {\cal H}_{L}(\alpha_x,\alpha_z) \, , \label{eq:pz}
\ea
where $\varepsilon_{\rm eq}$ and $P_{\rm eq}$ are the isotropic equilibrium energy density and pressure, respectively, and the special functions ${\cal R}$, ${\cal H}_{Tx}$, ${\cal H}_{Ty}$, and ${\cal H}_{L}$ are defined in App.~\ref{app:special-functions}.  The derivatives
\ba
\theta_u&\equiv&\partial_\mu u^\mu\,, \\
\theta_x&\equiv&\partial_\mu X^\mu\,, \\
D_u &\equiv& u^\mu D_\mu\,, \\
D_x &\equiv& X^\mu D_\mu\,,
\ea
are defined in the 1+1d case in App.~\ref{app:derivatives} and $\sigma_i$ are the diagonal projections of the velocity stress tensor $\sigma_{\mu\nu} \equiv \nabla_{\!\langle\mu}u_{\nu\rangle}$
\ba
\sigma_x &\equiv& X^\mu \sigma_{\mu\nu} X^\nu = \frac{\cosh\theta_\perp}{\tau}+\frac{\sinh\theta_\perp}{r} - \frac{2 \theta_u}{3} \, , \\
\sigma_y &\equiv& Y^\mu \sigma_{\mu\nu} Y^\nu = \frac{\theta_u}{3} - \frac{\sinh\theta_\perp}{r} \, .
\label{eq:xi-alpha}
\ea

The fifth equation necessary is obtained by requiring energy conservation.  In the conformal case, this constraint implies that the first moment of the collisional kernel must vanish and results in the dynamical Landau-Matching condition
\be
T = \lambda R^{1/4}(\alpha_x,\alpha_y) \, .
\label{eq:landau-matching}
\ee
In the next section, we will describe how to extend this to the non-conformal case.
Finally, we note that the $\alpha_i$'s obey the following constraint\,\footnote{For a non-conformal system, the right hand side of Eq.~(\ref{eq:constraint}) would instead be $3(1+\Phi)$.}
\be
\frac{1}{\alpha_x^2}+\frac{1}{\alpha_y^2}+\frac{1}{\alpha_z^2} = 3 \, ,
\label{eq:constraint}
\ee
which can be used to determine $\alpha_y$ as a function of $\alpha_x$ and $\alpha_z$
\be
\alpha_y(\alpha_x,\alpha_z)=\frac{\alpha_x\alpha_z}{\sqrt{3\alpha^2_x\alpha^2_z-\alpha^2_x-\alpha^2_z}} \, .
\label{eq:ay}
\ee

\section{Lattice-based equation of state}
\label{sec:eos}

The dynamical equations presented in the previous section were obtained \cite{Tinti:2013vba,Nopoush:2014qba} in the conformal (massless) limit.  In this case, there is no fundamental scale and one has
\ba
P_{\rm eq}\rightarrow P_{\rm ideal}\, , \nonumber \\
\varepsilon_{\rm eq}\rightarrow \varepsilon_{\rm ideal}\, , \label{eq:ideal-eso0}
\ea
with
\be
\varepsilon_{\rm ideal}=3P_{\rm ideal} \, .
\label{eq:ideal-eso1}
\ee
For the quark-gluon plasma the ideal (Stefan-Boltzmann) limit of the pressure is 
\be
P_{\rm ideal}=P_{\rm SB}=\frac{\pi^2}{45}T^4\left(N_c^2-1+\frac{7}{4}N_c N_f\right) ,
\ee
where $N_c$ and $N_f$ are number of colors and quark flavors, respectively. 

In practice, however, interactions induce corrections to the ideal EoS and also result in the running of the strong coupling constant which breaks conformal invariance.  At low temperatures, the quark-gluon plasma behaves non-conformally and one must use lattice QCD simulations to determine the EoS.  Herein, we employ an analytic parameterization of lattice data for the QCD interaction measure, $I(T) = \varepsilon_{\rm eq} - 3 P_{\rm eq}$, taken from the Wuppertal-Budapest collaboration \cite{Borsanyi:2010cj}
\be
\frac{I(T)}{T^4}=\bigg[\frac{h_0}{1+\alpha t^2}+\frac{f_0\big[\tanh(f_1t+f_2)+1\big]}{1+g_1t+g_2t^2}\bigg]\exp\!\Big(\!-\!\frac{h_1}{t}-\frac{h_2}{t^2}\Big)\, .
\label{eq:I-func}
\ee
with $t\equiv T/(0.2 \; \rm GeV)$. For $n_f=2+1$ (2 light and one heavy quarks) the parameters are $h_0=0.1396$, $h_1=-0.1800$, $h_2=0.0350$, $f_0=2.76$, $f_1=6.79$, $f_2=-5.29$, $g_1=-0.47$, $g_2=1.04$, and $\alpha=0.01$.\footnote{In the original parameterization presented in Ref.~\cite{Borsanyi:2010cj} the authors used $\alpha=0$, however, choosing $\alpha=0$ gives the wrong high temperature limit.  We introduce a small $\alpha > 0$ which does not affect the parameterization in the vicinity of the phase transition, where the original fitting was performed, but guarantees that the pressure approaches the Stefan-Boltzmann limit in the high-temperature limit.}

\begin{figure}[t]
\hspace{-6mm}
\includegraphics[width=.46\linewidth]{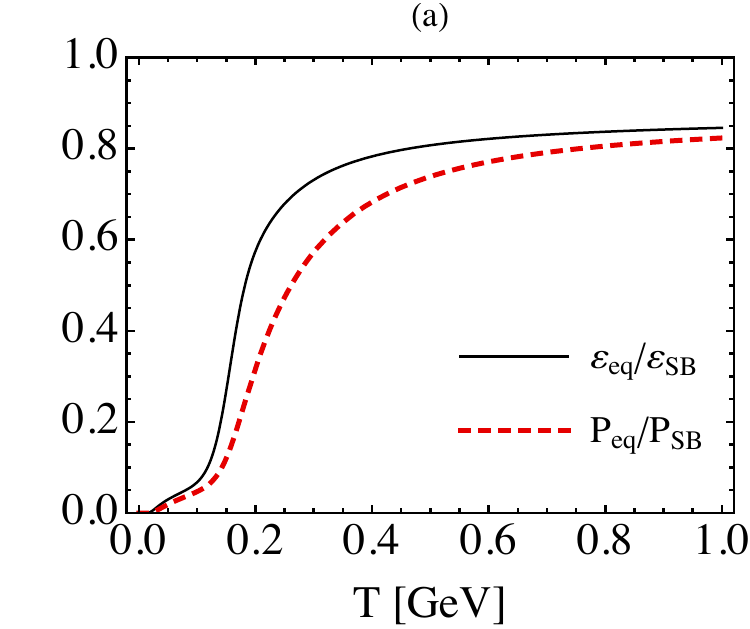}
\hspace{4mm}
\includegraphics[width=.49\linewidth]{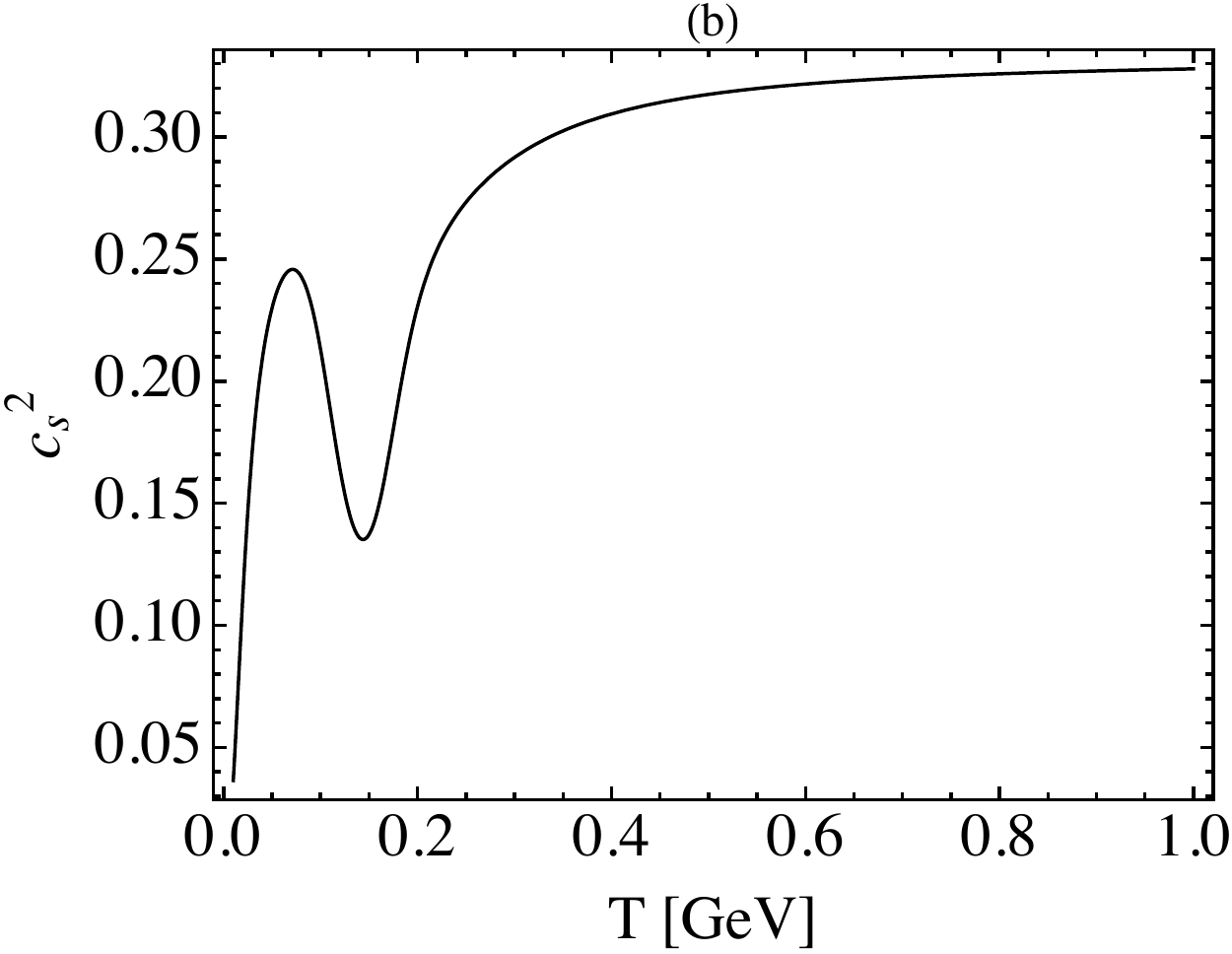}
\caption{The left panel (a) shows the equilibrium energy density and pressure obtained from Eq.~(\ref{eq:I-func}), scaled by their ideal Stefan-Boltzmann limit. The right panel (b) shows the speed of sound $(c_s^2= \partial P_{\rm eq}/\partial\varepsilon_{\rm eq})$ as a function of temperature.}
\label{fig:eos}
\end{figure}

The pressure can be obtained from an integral of the interaction measure 
\be
\frac{P_{\rm eq}(T)}{T^4}=\int_0^T \frac{dT}{T}\frac{I(T)}{T^4} \, ,
\label{eq:P-func}
\ee
where we have assumed $P(T=0)=0$.  Having $P_{\rm eq}(T)$, one can obtain the energy density $\varepsilon_{\rm eq}$ via
\be
\varepsilon_{\rm eq} = 3 P_{\rm eq}(T) + I(T)\, .
\label{eq:realistic-eos}
\ee
Note also, that one can construct the inverse function $T(\varepsilon)$ straightforwardly.

In the limit $T\rightarrow\infty$, the system tends to the ideal limit (\ref{eq:ideal-eso1}) as expected. The temperature dependence of the resulting isotropic energy density, pressure, and speed of sound squared ($c_s^2 = \partial P_{\rm eq}/\partial\varepsilon_{\rm eq}$) are shown in the two panels of Fig.~\ref{fig:eos}.  In what follows, the temperature dependence of the lattice-parameterized $\varepsilon_{\rm eq}$ and $P_{\rm eq}$ are used when computing the anisotropic energy density and pressures via Eqs.~(\ref{eq:ed})-(\ref{eq:pz}) and the inverse function $T(\varepsilon)$ is used where $T$ appears in Eqs.~(\ref{eq:equation-1})-(\ref{eq:equation-3}).
In addition to using the self-consistent effective temperature and lattice-based EoS for $\varepsilon_{\rm eq}$ and $P_{\rm eq}$, one must also specify how the relaxation time is determined.  In the relaxation time approximation with a general EoS, one has $\tau_{\rm eq} = 5 \eta/4 P_{\rm eq}$.  In the conformal limit, one can use $4 P_{\rm eq} = \varepsilon_{\rm eq} + P_{\rm eq} = T s$ to rewrite this as $\tau_{\rm eq,\,conformal} = 5 \bar\eta/T$.  For a general EoS, if one works with fixed $\bar\eta=\eta/s$, one has instead \mbox{$\tau_{\rm eq} = 5 \bar\eta \, (1 + \varepsilon_{\rm eq}/P_{\rm eq})/4T$}.

Before proceeding, we note that there is somewhat of a inconsistency in our prescription for implementing the EoS in anisotropic hydrodynamics.  This stems from the fact that the factorization (\ref{eq:ed})-(\ref{eq:pz}), which occurs in the conformal case, no longer holds in the case of a non-conformal (massive) gas \cite{Martinez:2012tu,Florkowski:2014sfa,Florkowski:2014bba}.  However, as we demonstrate in App.~\ref{app:factorization}, one finds that for $0.1 \lesssim P_L/P_T \lesssim 10$ and masses $m/T \lesssim 1$ the factorization of the thermodynamic variables is accurate to $\lesssim$ 5\%, with the largest corrections being seen for a strongly prolate plasma ($P_L \gg P_T$).  Although we have no precise knowledge of the effective degrees of freedom and their masses in a QGP in the temperature range considered here, this raises some hope that the conformal factorization approach used herein is a reasonable approximation.  However, even with this understanding, there is another complication if one breaks the conformal symmetry since, in this case, one cannot naively assume that the parameter $\Phi$ appearing in the general LO ansatz for the one-particle distribution function (\ref{eq:aniso-tensor1}) is zero \cite{Martinez:2012tu,Tinti:2013vba,Nopoush:2014qba}.  In a forthcoming paper, we study an alternative prescription for imposing the EoS which employs a quasiparticle model with a temperature-dependent quasiparticle mass that is tuned to reproduce the lattice EoS \cite{Mubarak:forth} as suggested recently in Ref.~\cite{Romatschke:2011qp}.  Our preliminary findings suggest that this alternative approach agrees quite well with the implementation chosen here for the evolution of the effective temperature and shear corrections, but that there may be important differences in the evolution of the bulk pressure correction, which, in the current approach, is related to the difference between $\alpha_x$ and $\alpha_z$.  We leave the comparison of these two methods to a future paper \cite{Mubarak:forth}.

\section{Anisotropic Freeze-out}
\label{sec:freeze-out}

We now turn to the question of hadronic freeze-out.  Our technique will be to perform ``anisotropic Cooper-Frye freeze-out'' using Eq.~(\ref{eq:pdf}) as the form for the one-particle distribution function.  This is different than the typical freeze-out prescription used in viscous hydrodynamics in which one takes into account the dissipative correction to the equilibrium distribution function only at linear order.  One immediate benefit of performing anisotropic freeze-out using Eq.~(\ref{eq:pdf}) is that, with this form, one is guaranteed that the one-particle distribution function is positive-definite at all space-time points in the plasma.

In practice, we start from the standard freeze-out integral
\be
N=\int_{\Sigma} d^3\Sigma_\mu j^\mu  \, .
\label{eq:particle-number1} 
\ee
In the integral above, $\Sigma$ is the three-dimensional freeze-out hypersurface defining the boundary of the four-dimensional volume occupied by the fluid, $d^3\Sigma^\mu$ is the surface normal vector, and $j^\mu$ is the particle four-current.  Due to the presence of momentum-space anisotropies, one cannot simply use the momentum scale $\lambda$ when defining the freeze-out hypersurface $\Sigma$.  Instead, one should use the energy density, from which one can obtain the effective freeze-out temperature $T_{\rm FO} = T_{\rm eff} = T(\varepsilon)$ using the realistic EoS described in the previous section. 

After identifying $\Sigma$, we use the following parametrization of the freeze-out hypersurface~\cite{Bozek:2009ty,Bozek:2009dw,Chojnacki:2011hb,Ryblewski:2015hea},\footnote{Different parametrizations have been used in the literature \cite{Kolb:2003dz,Heinz:2004qz,Muronga:2004sf,Baier:2006gy}. The parametrization used here has the advantage that the function $d(\zeta,\phi,\theta)$ is single-valued for most (but not all) freeze-out surfaces, including the typical one shown in Fig.~\ref{fig:vacompfreeze10} for which other parameterizations, e.g.~using the freeze-out proper time $\tau_\mathrm{FO}(r)$, are multivalued.}
\ba
  t &=& \big(\tau_0 + d(\zeta,\phi,\theta) \sin\theta \sin\zeta \big)\cosh\!\left[\frac{d(\zeta,\phi,\theta) \cos\theta}{\Lambda}\right] , \nonumber \\
  x &=& d(\zeta,\phi,\theta) \sin\theta \cos\zeta \cos\phi \, , \nonumber \\
  y &=& d(\zeta,\phi,\theta) \sin\theta \cos\zeta \sin\phi \, , \nonumber \\
  z &=& \big(\tau_0 + d(\zeta,\phi,\theta) \sin\theta \sin\zeta \big)\sinh\!\left[\frac{d(\zeta,\phi,\theta) \cos\theta}{\Lambda}\right] .
 \label{eq:minko-par}
\ea
Based on this one finds the following expressions for the longitudinal proper time $\tau$, longitudinal rapidity $\varsigma$, and the transverse ($r$ and $\phi$) coordinates of $\Sigma$
\begin{eqnarray}
 \tau &=& \tau_0 + d(\zeta,\phi,\theta) \sin\theta \sin\zeta \nonumber  \, ,
 \nonumber   \\
 r &=&  d(\zeta,\phi,\theta) \sin\theta \cos\zeta  \, ,\nonumber \\
 \phi &=&\phi\, , \nonumber  \\
 \varsigma &=&  \frac{d(\zeta,\phi,\theta) \cos\theta}{\Lambda}  \, .
  \label{eq:milne-par}
\end{eqnarray}
The function $d(\zeta,\phi,\theta)$ is by construction the distance of the points on freeze-out hypersurface to the point $(\tau_0,0,0,0)$ in $(\tau,r,\phi,\Lambda\varsigma)$-system of coordinates ($\tau_0$ being initial proper time) as can be seen from Eq.~(\ref{eq:milne-par}). The freeze-out surface variables $\theta$ and $\zeta$ are the polar and azimuthal coordinates in this coordinate system.  The length scale $\Lambda$ is introduced for dimensional reasons and final results are independent of this quantity.  The normal vector to the hypersurface is constructed in the usual manner by taking derivatives of the orthogonal basis coordinates (\ref{eq:minko-par}) with respect to the relevant parameters $\zeta$, $\phi$, and $\theta$
\be
d^3\Sigma_\mu = \epsilon_{\mu\alpha\beta\gamma} \frac{\partial x^\alpha}{\partial \zeta}\frac{\partial x^\beta}{\partial \phi}\frac{\partial x^\gamma}{\partial \theta}d\zeta d\phi d\theta\, ,
\label{eq:dSigma}
\ee
where $\epsilon_{\mu\alpha\beta\gamma}$ is the four-dimensional Levi-Civita symbol. 

To proceed we use the kinetic definition of $j^\mu$
\be
j^\mu=\int d\chi \, p^\mu f(x,t)\, .
\label{eq:4current-int}
\ee
For energy densities below the QCD phase transition temperature (energy density), it is appropriate to describe the system as a gas of hadrons; therefore, $d\chi$ translates to 
\be
d\chi\equiv \sum\limits_i(2s_i+1)(2g_i+1)\frac{d^4p}{(2\pi)^3}\delta(p^\mu p_\mu-m_i^2)2\Theta(p^0) \, ,
\label{eq:dchi}
\ee
where $s_i$ and $g_i$ are the spin and isospin degeneracies of the hadron and $m_i$ is the hadron mass. 
Putting everything together one has 
\be
N=\sum\limits_i\int \! d^3p \frac{1}{\sqrt{m^2_i+p^2}}\bigg(p^0\frac{dN}{d^3p}\bigg)_i \, ,
\label{eq:N}
\ee
with 
\be
\bigg(p^0\frac{dN}{d^3p}\bigg)_{\!\!i}=\frac{{\cal N}_i}{(2\pi)^3}\int \! f_i(x,p) \, p^\mu d^3\Sigma_\mu \, ,
\label{eq:dNdp3}
\ee
where ${\cal N}_i\equiv(2s_i+1)(2g_i+1)$ is the degeneracy factor and $f_i$ is the distribution function for the particle species $i$ taking into account the appropriate quantum statistics. 

Parameterizing the particle momentum in the lab frame as 
\ba
p^\mu\equiv(m_\perp \cosh y,p_\perp \cos\varphi,p_\perp \sin\varphi,m_\perp \sinh y)\, ,
\label{eq:ptl-mom}
\ea
one finds
\ba
p \cdot u &=& m_\perp \cosh (\theta _\perp) \cosh (y-\varsigma)-p_\perp \sinh (\theta _\perp)
   \cos (\phi -\varphi )\, ,\nonumber
    \\
p \cdot X &=& m_\perp \sinh (\theta _\perp) \cosh (y-\varsigma)-p_\perp \cosh (\theta _\perp)
   \cos (\phi -\varphi )
\, ,\nonumber \\
p \cdot Y &=& p_\perp \sin (\phi -\varphi )
\, ,\nonumber \\
p \cdot Z &=& -m_\perp \sinh (y-\varsigma)
\, ,
\label{eq:pdotuxyz}
\ea
where $m_\perp = \sqrt{p_\perp^2+m^2}$, $y = \tanh^{-1}(p^z/p^0)$ is the particle's rapidity, and $\varphi$ is the particle's azimuthal angle.
In order to set up the distribution function, having $p^\mu$ defined in Eq.~(\ref{eq:ptl-mom}), one can use (\ref{eq:aniso-tensor1}),  (\ref{eq:aniso-tensor2}), (\ref{eq:pdf}), and  (\ref{eq:pdotuxyz}) to find
\ba
p^\mu \Xi_{\mu\nu} p^\nu &=& 
\Big[m_\perp \cosh \theta _\perp \cosh
   (y-\varsigma)-p_\perp \sinh \theta _\perp \cos (\phi -\varphi )\Big]^2
   \nonumber \\ 
   &+& \xi _x\,\Big[m_\perp \sinh \theta _\perp \cosh (y-\varsigma)-p_\perp \cosh \theta
   _\perp \cos (\phi -\varphi )\Big]^2
   \nonumber \\
   &+& \xi_z\, m_\perp^2  \sinh^2(y-\varsigma)
   \nonumber \\ 
   &+&\xi_y\,\, p_\perp^2  \sin^2(\phi -\varphi )\,.
\label{eq:pxip}
\ea
Expanding Eq.~(\ref{eq:dSigma}) for $x^\mu=(t,x,y,z)$ and contracting with $p^\mu$, one obtains
\ba 
p^\mu d^3\Sigma_\mu  &=& \frac{\tau}{\Lambda}\sin\theta d^2 \Bigg[ p_\perp  \sin (\phi -\varphi ) \frac{\partial d}{\partial\phi} \notag \\ 
&+&\frac{\Lambda}{\tau}  m_\perp\cos\zeta \sin\theta \sinh(y-\varsigma) \bigg(d \cos\theta+\frac{\partial d}{\partial\theta}\sin\theta\bigg)\, \notag \\
&+&\cos\zeta \sin\theta\bigg(p_\perp \cos\zeta \cos (\phi -\varphi )+m_\perp \sin\zeta \cosh(y-\varsigma) \bigg)\!\bigg(d\sin\theta-\frac{\partial d}{\partial \theta}\cos\theta\bigg) \, \notag \\
&+&\cos\zeta\frac{\partial d}{\partial\zeta} \Big(p_\perp \sin \zeta \cos (\phi-\varphi) -m_\perp \cos\zeta \cosh(y-\varsigma)\Big) \Bigg]d\zeta d\phi d\theta\, ,
\label{eq:pdSigma}
\ea 
where $d\equiv d(\zeta ,\phi ,\theta )$.

\subsection{Boost-invariant and cylindrically-symmetric system}

If the system is cylindrically symmetric, one has
\be
 d(\zeta,\phi,\theta)\rightarrow d(\zeta,\theta)\, .
\label{eq:d}
\ee
If, in addition, the system is boost invariant along the beam-line direction, the hypersurface in ($\tau r\phi\varsigma$)-space is constant with respect to the longitudinal rapidity $\varsigma$.  As a consequence, $d(\zeta,\theta)\sin\theta$ (which is the projection of $d(\zeta,\theta)$ normal to the $\varsigma$-axis) should be constant, i.e.
\be
 d(\zeta,\theta)\sin\theta={\rm constant}\,. 
 \label{eq:d2}
\ee
Using this and taking the constant to be the value of the function at a typical longitudinal rapidity, i.e. $\varsigma=0$ ($\theta=\pi/2$), one finds
\ba
 d(\zeta,\theta) &=& \frac{d(\zeta)}{\sin\theta}\, , \notag \\
\frac{\partial d(\zeta,\theta)}{\partial \theta}&=& -d(\zeta) \rm \frac{cot\theta}{\sin\theta}\,,
\label{eq:d3}
\ea
where $d(\zeta)\equiv d(\zeta,\theta=\pi/2)$. Using the above simplifications, one finds the boost-invariant cylindrically-symmetric form of $p^\mu d^3\Sigma_\mu$
\ba
p^\mu d^3\Sigma_\mu = \frac{\tau}{\Lambda}  d(\zeta )^2 \cos \zeta  \csc ^2\theta \Bigg[d'(\zeta ) \Big(p_\perp \sin \zeta \cos (\phi -\varphi )-m_\perp \cos \zeta  \cosh (y-\varsigma)\Big) \notag \\ 
+d(\zeta ) \Big(p_\perp \cos \zeta \cos (\phi -\varphi )+m_\perp \sin\zeta  \cosh (y-\varsigma)\Big)\Bigg]d\zeta d\phi d\theta\, .
\label{eq:pdSigma2}
\ea

\section{Numerical Results}
\label{sec:results}

We now turn to our numerical results.  We present comparisons of results obtained using the dynamical equations of anisotropic hydrodynamics presented in Sec.~\ref{sec:equations} and the second-order viscous hydrodynamics equations from Denicol et al. \cite{Denicol:2012cn}.\footnote{For the smooth initial conditions considered herein, the vorticity is zero at all times.  We also set the transport coefficient $\tau_{\pi\pi}$ to zero since this has been done in almost all other implementations to date (see, however, \cite{Ryu:2015vwa}). As a result, we drop the last two terms in Eq.~(\ref{eq:vhyd3-1}).}  For anisotropic hydrodynamics, we use the freeze-out method detailed in Sec.~\ref{sec:freeze-out} and, for second-order viscous hydrodynamics, we use the standard Grad-14 ansatz for the dissipative correction to the one-particle distribution function.  We present the full details of both the viscous hydrodynamics equations and the freeze-out prescription used for our comparisons in App.~\ref{app:vhydro}.  For both anisotropic and second-order viscous hydrodynamics we use the lattice-parameterization of the EoS presented in Sec.~\ref{sec:eos}.  Both codes were tested by comparing the evolution of the system initialized with a Gubser temperature- and flow-profile and then comparing with the exact solution appropriate to each framework.  In both cases, using the lattice spacing, temporal step size, etc. specified below, we were able to reproduce the corresponding exact Gubser solution to very high accuracy at all times.  We present the details of our code tests in App.~\ref{app:tests}.  The code used to produce all figures in this text is publicly available~\cite{ahydrorepo}. 

\begin{figure}[t]
\includegraphics[width=0.85\linewidth]{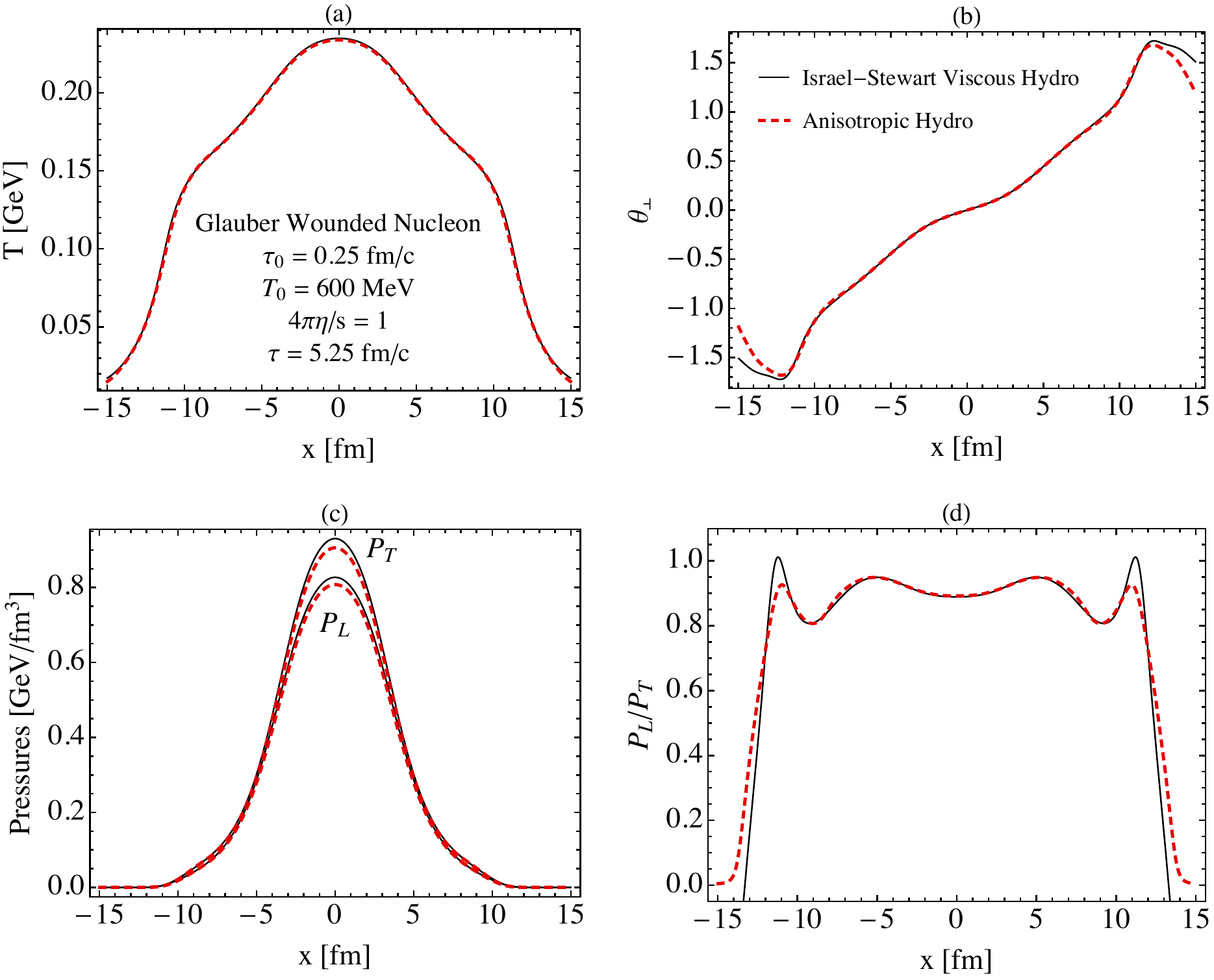}
\caption{Comparison of the numerical 1+1d solution to the Israel-Stewart and anisotropic hydrodynamics equations at $\tau = 5.25$ fm/c.  The shear viscosity to entropy density ratio was taken to be $4\pi\eta/s = 1$.  The four panels show (a) the effective temperature, (b) the transverse flow rapidity ($\theta_\perp$), (c) the transverse and longitudinal pressures, and (d) the ratio of the LRF longitudinal and transverse pressures. }
\label{fig:vacomp1}
\end{figure}

\begin{figure}[t]
\includegraphics[width=1\linewidth]{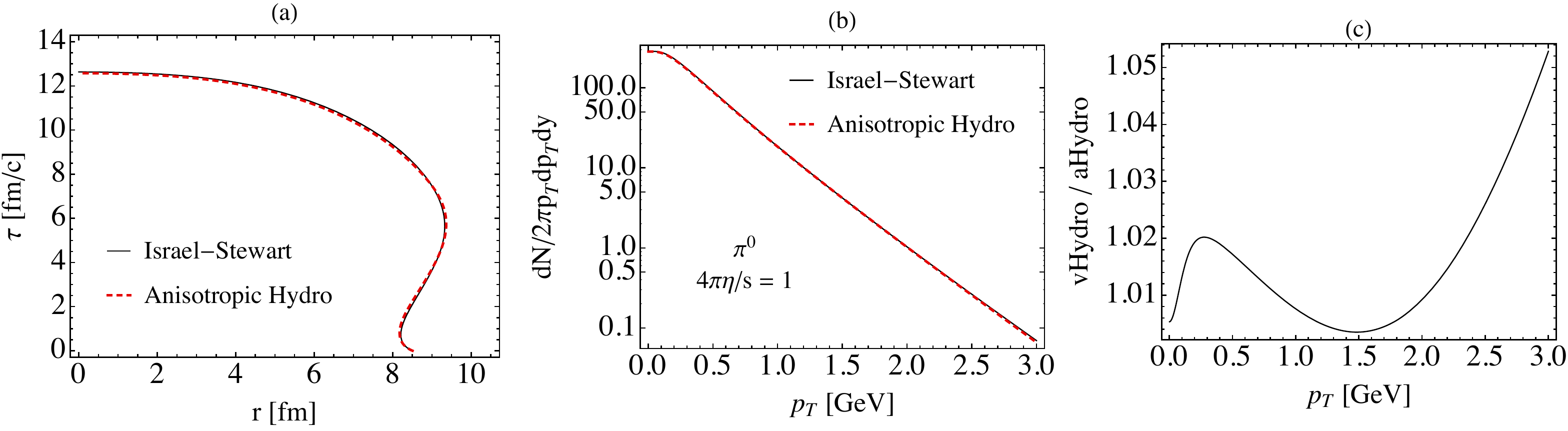}
\caption{Comparison of the numerical 1+1d solution to the Israel-Stewart and anisotropic hydrodynamics equations for the same conditions as shown in Fig.~\ref{fig:vacomp1}.  The three panels show (a) the freeze-out hypersurfaces, (b) resulting neutral pion spectra, and (c) the ratio of the neutral pion spectra obtained using Israel-Stewart and anisotropic hydrodynamics.}
\label{fig:vacompfreeze1}
\end{figure}

\subsection{Nucleus-nucleus collisions}

For all results presented in this section we use the Glauber wounded-nucleon overlap to set the initial energy density.  As our test case we consider Pb-Pb collisions with a center of mass energy of 2.76 GeV/nucleon.  We take the inelastic nucleon-nucleon scattering cross-section to be $\sigma_{NN} = $ 62 mb. We use 300 points in the radial direction with a lattice spacing of $\Delta r = 0.05$ fm and temporal step size of $\Delta \tau = $ 0.01 fm/c.  We use fourth-order Runge-Kutta integration for the temporal updates and fourth-order centered differences for the evaluation of all spatial derivatives.\footnote{Since the initial conditions considered herein are smooth, naive centered differences generally suffice.} Unless otherwise indicated, we take the central initial temperature to be $T_0 = 600$ MeV at $\tau_0 = 0.25$ fm/c and assume that the system is initially isotropic, i.e. $\alpha_x(\tau_0) = \alpha_z(\tau_0) = 1$ for anisotropic hydrodynamics and $\pi^{\mu\nu}(\tau_0)=0$ for second-order viscous hydrodynamics.  We take the freeze-out temperature to be $T_{\rm eff} = T_{\rm FO} = $ 150 MeV in all cases.

\subsubsection{Hydrodynamic evolution and spectra}

In Fig.~\ref{fig:vacomp1} we present a comparison of the numerical 1+1d solution to the Israel-Stewart and anisotropic hydrodynamics equations for the case $4\pi\eta/s = 1$.  As can be seen from Figs.~\ref{fig:vacomp1}a and \ref{fig:vacomp1}b, for this small value of the shear viscosity to entropy ratio, the two methods agree quite well, with only small differences seen in both the temperature and transverse rapidity profiles at the time shown.  At very early times, the differences are somewhat larger, but by $\tau = 5.25$ fm/c the agreement is quite good.  Note that the bumps located at $r\simeq10$ are a reflection of evolution near the softest point in the EoS and both the standard viscous hydrodynamic method for imposing the EoS and the anisotropic hydrodynamics method implemented herein seem to give comparable results.  Although we don't show it here, we also performed tests for $4 \pi \eta/s = 0.1$ and found that anisotropic hydrodynamics and Israel-Stewart second-order hydrodynamics give virtually indistinguishable results in this case.

In Figs.~\ref{fig:vacomp1}c and \ref{fig:vacomp1}d we present a comparison of the results obtained for the local rest frame (LRF) transverse and longitudinal pressures, again for the case of $4\pi\eta/s = 1$.  For anisotropic hydrodynamics, we compute $P_x$ using Eq.~(\ref{eq:px}) which corresponds to the radial pressure since, using the vectors specified in Eq.~(\ref{eq:expvectorbasis}), one sees that $X^\mu$ maps to the radial direction in the LRF.  In the case of second-order viscous hydrodynamics we identify $P_T = P_{\rm eq} + \pi^r_r$ and $P_L = P_{\rm eq} + \pi^\varsigma_\varsigma$, again in the LRF.  As these panels demonstrate, for $4\pi\eta/s = 1$ the pressures obtained are similar using both methods with the the maximum difference for $r \lesssim 10$ fm being less than approximately 2\%.  At very large $r$ we see larger differences, with the longitudinal pressure predicted by the Israel-Stewart equations becoming negative.  However, the temperature where this occurs is quite small and far below the freeze-out temperature.  We do note that one finds that anisotropic hydrodynamics predicts that the system is generally closer to isotropy than Israel-Stewart hydrodynamics.  As we will see below, the smaller pressure anisotropy effectively reduces the shear stress contribution to the transverse pressure, reducing the buildup of radial flow and leading to softer particle spectra from anisotropic hydrodynamics.

Next, we turn to Fig.~\ref{fig:vacompfreeze1}.  In this figure, we present comparisons of the freeze-out hypersurfaces (Fig.~\ref{fig:vacompfreeze1}a), the resulting neutral pion spectrum (Fig.~\ref{fig:vacompfreeze1}b), and the ratio of the neutral pion spectra obtained using Israel-Stewart and anisotropic hydrodynamics (Fig.~\ref{fig:vacompfreeze1}c).  As these panels demonstrate one finds that the freeze-out hypersurfaces and resulting particle spectra are quite similar for $4\pi\eta/s = 1$ which should not be overly surprising.  As Fig.~\ref{fig:vacompfreeze1}c demonstrates the maximal correction for $p_T <$ 3 GeV is approximately 5\%, with second-order viscous hydrodynamics predicting a harder distribution with a slightly higher mean-$p_T$.

\begin{figure}[ht!]
\includegraphics[width=0.85\linewidth]{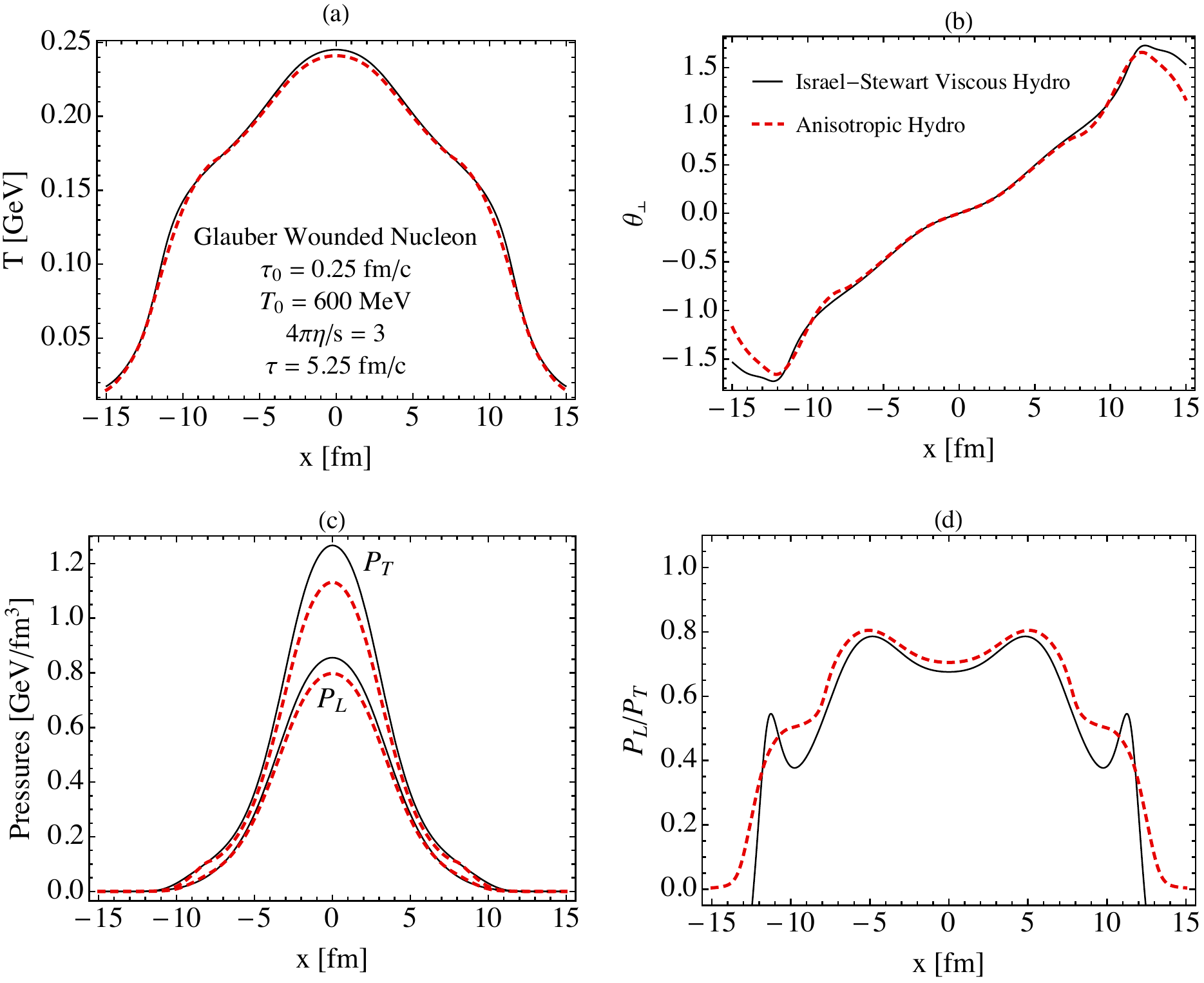}
\caption{Comparison of the numerical 1+1d solution to the Israel-Stewart and anisotropic hydrodynamics equations.  The shear viscosity to entropy density ratio was taken to be $4\pi\eta/s = 3$.  The quantities shown in the panels and the initial conditions are the same as in Fig.~\ref{fig:vacomp1}.}
\label{fig:vacomp3}
\end{figure}

\begin{figure}[ht!]
\includegraphics[width=1\linewidth]{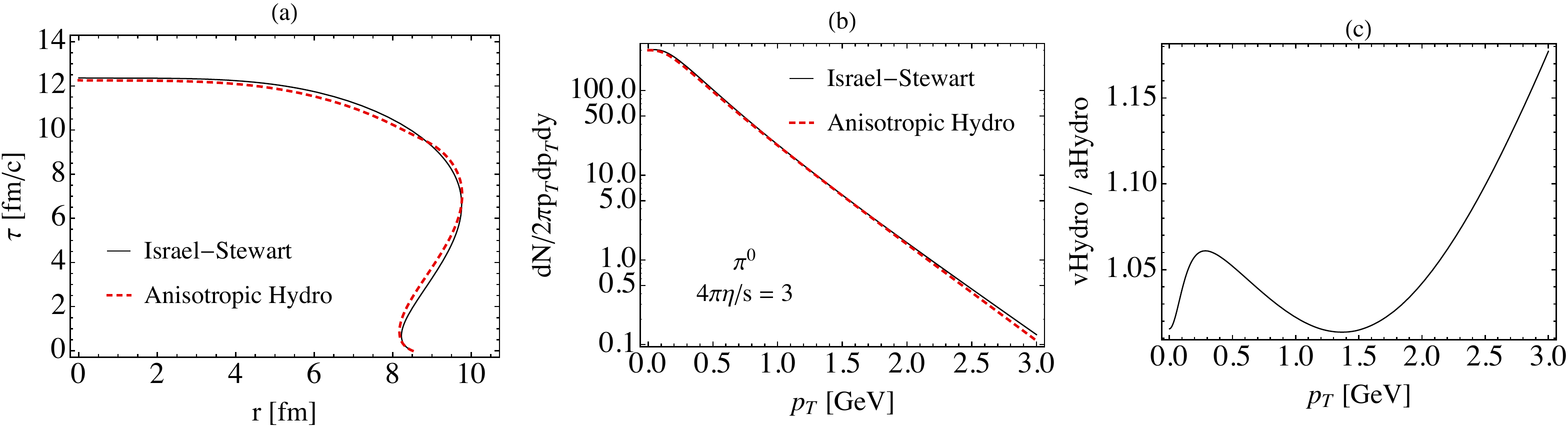}
\caption{Comparison of the numerical 1+1d solution to the Israel-Stewart and anisotropic hydrodynamics equations for the same conditions as shown in Fig.~\ref{fig:vacomp3}.  The quantities shown in the panels are the same as in Fig.~\ref{fig:vacomp1}.}
\label{fig:vacompfreeze3}
\end{figure}

\begin{figure}[t]
\includegraphics[width=0.97\linewidth]{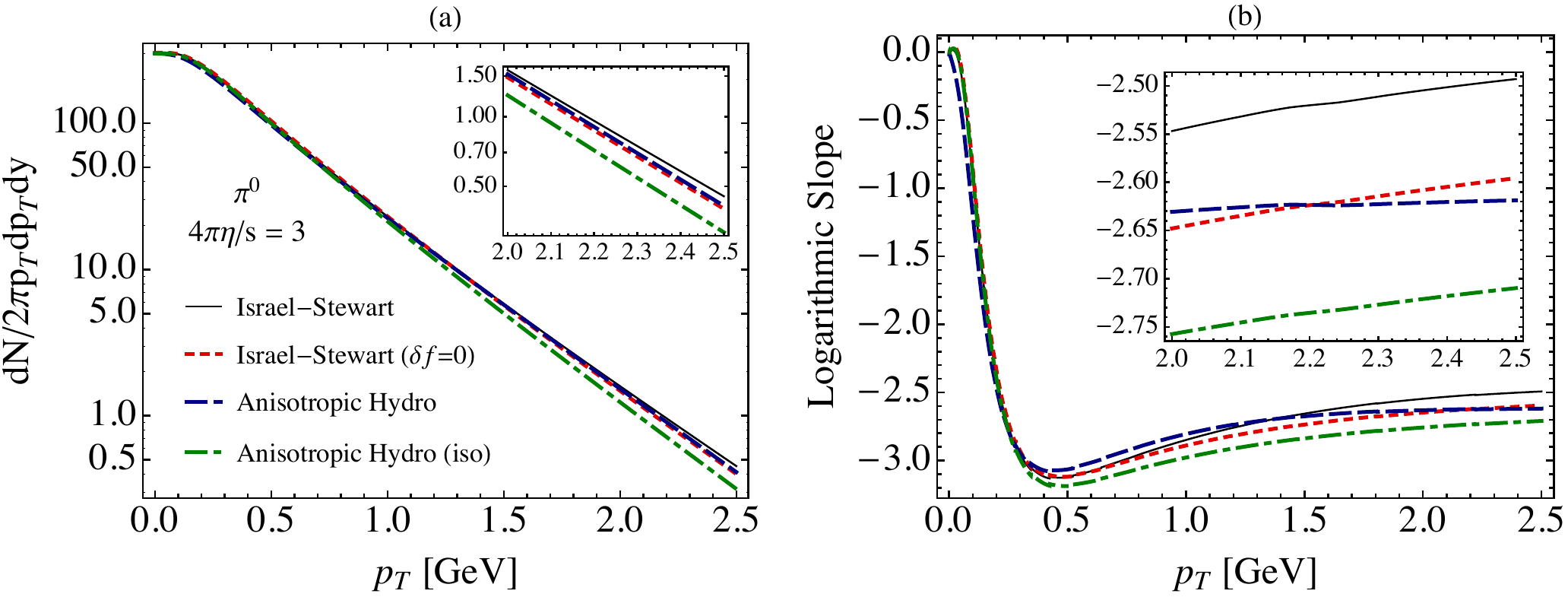}
\caption{Comparison of the $\pi^0$ spectra coming from Israel-Stewart with and without ($\delta f=0$) viscous corrections to the distribution function and anisotropic hydrodynamics with a complete treatment and assuming isotropic freeze-out using the effective temperature.  In this figure, we use the same conditions and parameters as shown in Fig.~\ref{fig:vacomp3}.  The two panels show (a) the resulting neutral pion spectra and (b) the logarithmic slope of the various curves.}
\label{fig:specComp1}
\end{figure}

\begin{figure}[t!]
\includegraphics[width=0.85\linewidth]{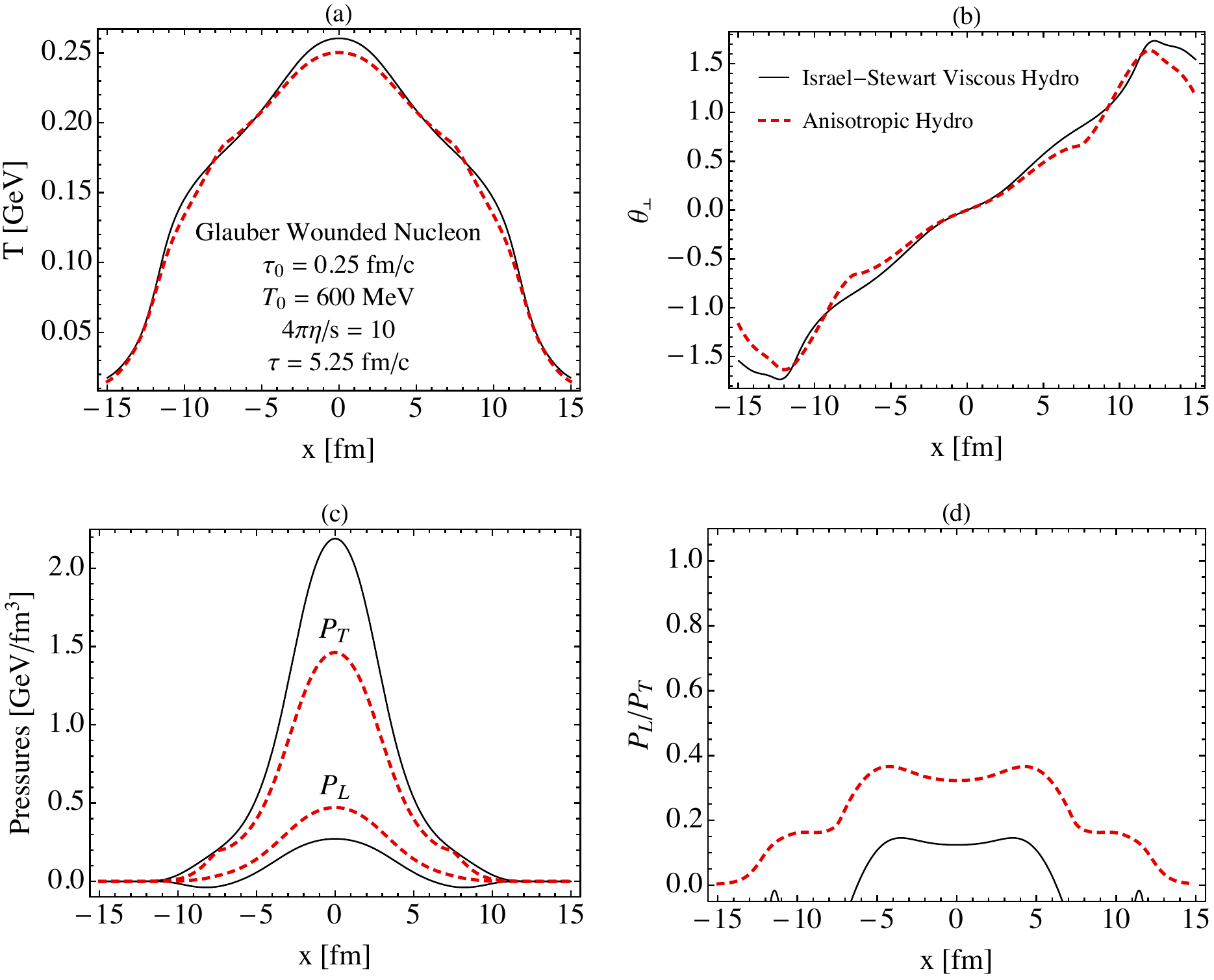}
\caption{Comparison of the numerical 1+1d solution to the Israel-Stewart and anisotropic hydrodynamics equations.  The shear viscosity to entropy density ratio was taken to be $4\pi\eta/s = 10$.  The quantities shown in the panels and the initial conditions are the same as in Fig.~\ref{fig:vacomp1}.}
\label{fig:vacomp10}
\end{figure}

\begin{figure}[t!]
\includegraphics[width=1\linewidth]{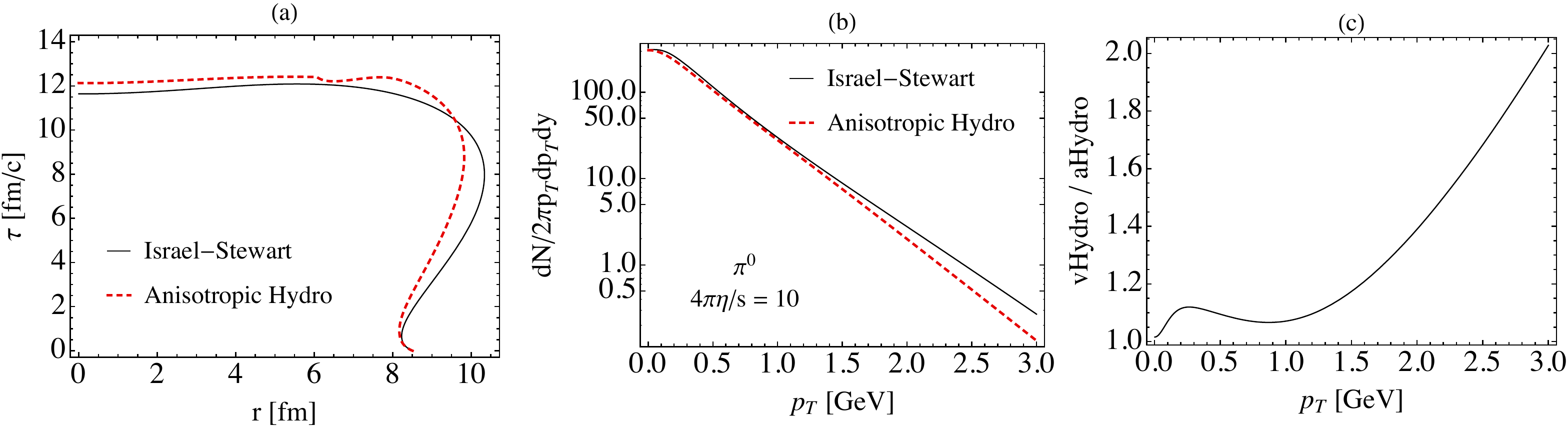}
\caption{Comparison of the numerical 1+1d solution to the Israel-Stewart and anisotropic hydrodynamics equations for the same conditions as shown in Fig.~\ref{fig:vacomp10}.  The quantities shown in the panels and the initial conditions were the same as in Fig.~\ref{fig:vacompfreeze1}.}
\label{fig:vacompfreeze10}
\end{figure}

In Figs.~\ref{fig:vacomp3} - \ref{fig:vacompfreeze10} we present similar plots for the cases $4\pi\eta/s = 3$ and $4\pi\eta/s = 10$.  As Fig.~\ref{fig:vacomp3} demonstrates, for $4\pi\eta/s = 3$ we find larger differences between anisotropic hydrodynamics and second-order viscous hydrodynamics, as could be expected a priori.  One sees that in both cases larger pressure anisotropies are generated than for $4\pi\eta/s = 1$ and the maximum difference in the pressure anisotropy approaching 70\% in the region $r<10$ fm.  The freeze-out hypersurface and neutral pion spectra shown in Fig.~\ref{fig:vacompfreeze3} also show larger differences, with the final neutral pion spectra having a maximum difference of approximately 18\% for $p_T < 3$ GeV.  Once again, we find that the neutral pion spectra predicted by second-order viscous hydrodynamics are harder than that predicted by anisotropic hydrodynamics.  

In order to further understand the differences seen in the predicted neutral pion spectra, in Fig.~\ref{fig:specComp1} we present the result of using different freeze-out prescriptions for both anisotropic hydrodynamics and second-order viscous hydrodynamics.  For anisotropic hydrodynamics, we show two cases:  (1) using the fully anisotropic distribution as in Eq.~(\ref{eq:pxip}) together with the local value of $\lambda$ to construct the local distribution via Eq.~(\ref{eq:pdf}) and (2) manually setting $\xi_x = \xi_y = \xi_z = 0$ in Eq.~(\ref{eq:pxip}) and setting $\lambda \rightarrow T$ which is the local effective isotropic temperature determined from the local energy density.  For viscous hydrodynamics, we also show two cases:  (1) Using the full Grad-14 form given in Eq.~(\ref{eq:vfreeze}) and (2) setting $\delta f = 0$, which corresponds to discarding the second term in square brackets in Eq.~(\ref{eq:vfreeze}).  As can be seen from Fig.~\ref{fig:specComp1}a, for both anisotropic hydrodynamics and second-order viscous hydrodynamics, the inclusion of the (anisotropic) dissipative correction to the one-particle distribution function results in a hardening of the particle spectra, however, for anisotropic hydrodynamics the correction is slightly smaller.  In order to separate the effect of an overall shift upwards in the spectra between the various cases, in Fig.~\ref{fig:specComp1}b we present the logarithmic slope of the spectra in all four cases.  This plot allows us to more directly see the hardening in each case.  Even without viscous corrections in the freeze-out distribution function (red dashed and green dash-dotted lines in Fig.~\ref{fig:specComp1}) the spectra obtained using Israel-Stewart viscous hydrodynamics are harder than those obtained using anisotropic hydrodynamics. This demonstrates that the hardening is primarily due to stronger radial flow, caused by larger transverse shear stress and reduced work done by the longitudinal pressure in Israel-Stewart hydrodynamics.   Viscous corrections at freeze-out further enhance the hardening of the spectra obtained using Israel-Stewart hydrodynamics.

Next we consider the case $4\pi\eta/s = 10$, which is shown in Figs.~\ref{fig:vacomp10} and \ref{fig:vacompfreeze10}.  As can be expected, we see larger differences between the two approaches in this case.  At the time shown, one sees that the longitudinal pressure predicted by second-order hydrodynamics is negative for all $r \gtrsim 6.5$ fm.  If one were to plot the longitudinal pressure predicted by second-order viscous hydrodynamics at much earlier times, one would find that the longitudinal pressure becomes negative in the entire simulation volume.  This is indicative of the breakdown of second-order viscous hydrodynamics in this case.  Not surprisingly, as can be seen from Fig.~\ref{fig:vacompfreeze10}, one sees large differences to both the freeze-out hypersurface and the final neutral pion spectra, with second-order viscous hydrodynamics predicting a significantly harder distribution and approximately 100\% more neutral pions at $p_T = $ 3 GeV.

\subsubsection{Dissipative particle production}

\begin{figure}[t!]
\includegraphics[width=0.85\linewidth]{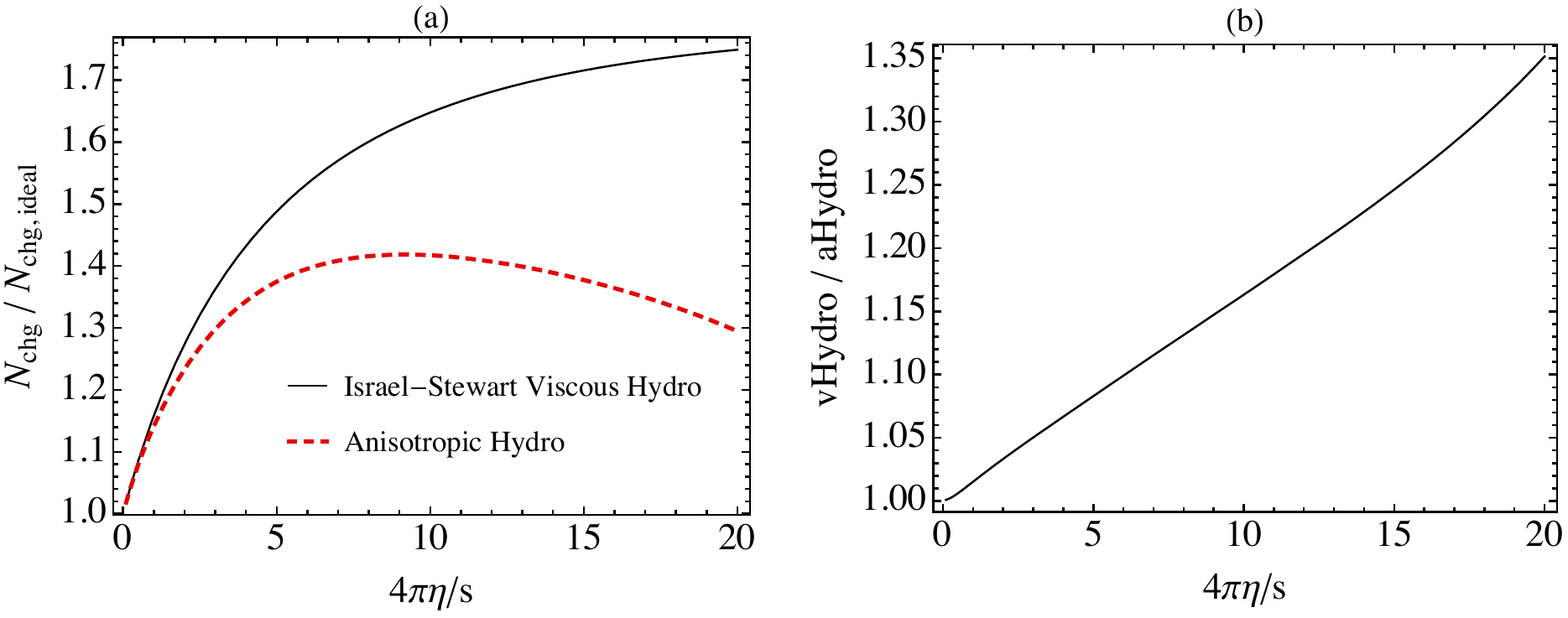}
\caption{The left panel (a) shows the number of charged particles scaled by the ideal hydrodynamics result as function of $4\pi\eta/s$ obtained using Israel-Stewart viscous hydrodynamics (black line) and anisotropic hydrodynamics (red dashed line).  The right panel (b) shows the ratio of the  Israel-Stewart viscous hydrodynamics result to the anisotropic hydrodynamics result.  The initial central temperature was taken to be $T_0 = $ 600 MeV.}
\label{fig:numberchg1}
\end{figure}

\begin{figure}[t!]
\includegraphics[width=0.85\linewidth]{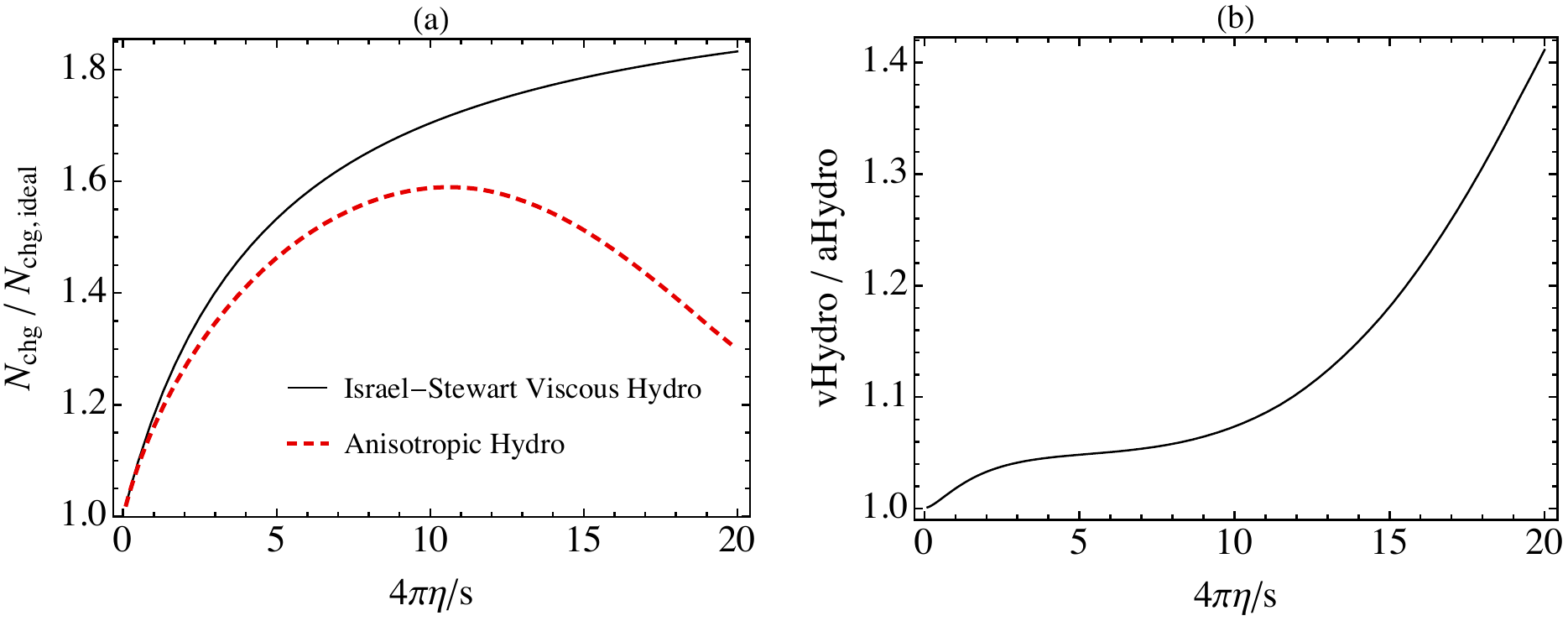}
\caption{Same as Fig.~\ref{fig:numberchg1}, except here we take the initial central temperature to be $T_0 = $ 500 MeV.}
\label{fig:numberchg2}
\end{figure}

As our last consideration in the context of Pb-Pb collisions, in Figs.~\ref{fig:numberchg1} and \ref{fig:numberchg2} we present the total number of charged particles, $N_{\rm chg}$, scaled by the ideal result, $N_{\rm chg, ideal}$, as a function of $4\pi\eta/s$.  For the purposes of this figure, we have included the production of charged pions, kaons, and protons.  In Fig.~\ref{fig:numberchg1} we took an initial central temperature of $T_0$ = 600 MeV at $\tau_0 = 0.25$ fm/c and in Fig.~\ref{fig:numberchg2} we took an initial central temperature of $T_0$ = 500 MeV at the same initial time.  As both figures demonstrate, one finds that second-order viscous hydrodynamics predicts that $N_{\rm chg}$ is a monotonically increasing function of $\eta/s$ whereas anisotropic hydrodynamics predicts that there is a maximum in charged particle production at $4 \pi \eta/s \sim$ 9 - 11 depending on the assumed initial temperature.  In panel (b) of both figures we show the ratio of the particle production predicted by second-order viscous hydrodynamics and that predicted by anisotropic hydrodynamics.  For $4\pi\eta/s\lesssim5$, both Figs.~\ref{fig:numberchg1} and \ref{fig:numberchg2} show that the difference in the total number of charged particles produced in Israel-Stewart and anisotropic hydrodynamics remains below 10\%. However, even a correction on the order of 10\% could have an important phenomenological impact.  Note that the non-monotonic behavior as a function of $\eta/s$ of particle production in anisotropic hydrodynamics is similar to results obtained in 0+1d anisotropic hydrodynamics \cite{Martinez:2012tu,Florkowski:2013lza,Bazow:2013ifa}, which can be straightforwardly understood as the vanishing of dissipative particle production in the free-streaming limit.

\subsection{Proton-Nucleus collisions}

We now consider the case of an asymmetric collision between a proton and a nucleus.  Of course, since our equations are boost invariant one cannot draw firm conclusions regarding comparisons with experimental results.  Our goal is to simply ascertain the magnitude of the differences one sees when using anisotropic hydrodynamics versus second-order viscous hydrodynamics for small systems.  Since the anisotropic hydrodynamics equations have been shown to better reproduce the spatiotemporal evolution for small systems subject to Gubser flow for all values of $\eta/s$ \cite{Nopoush:2014qba}, we believe that the anisotropic hydrodynamics equations used herein should also provide a more faithful reproduction of the bulk evolution for strong dynamically generated 1+1d flows in small collisional systems at high energies.

\begin{figure}[t!]
\includegraphics[width=0.85\linewidth]{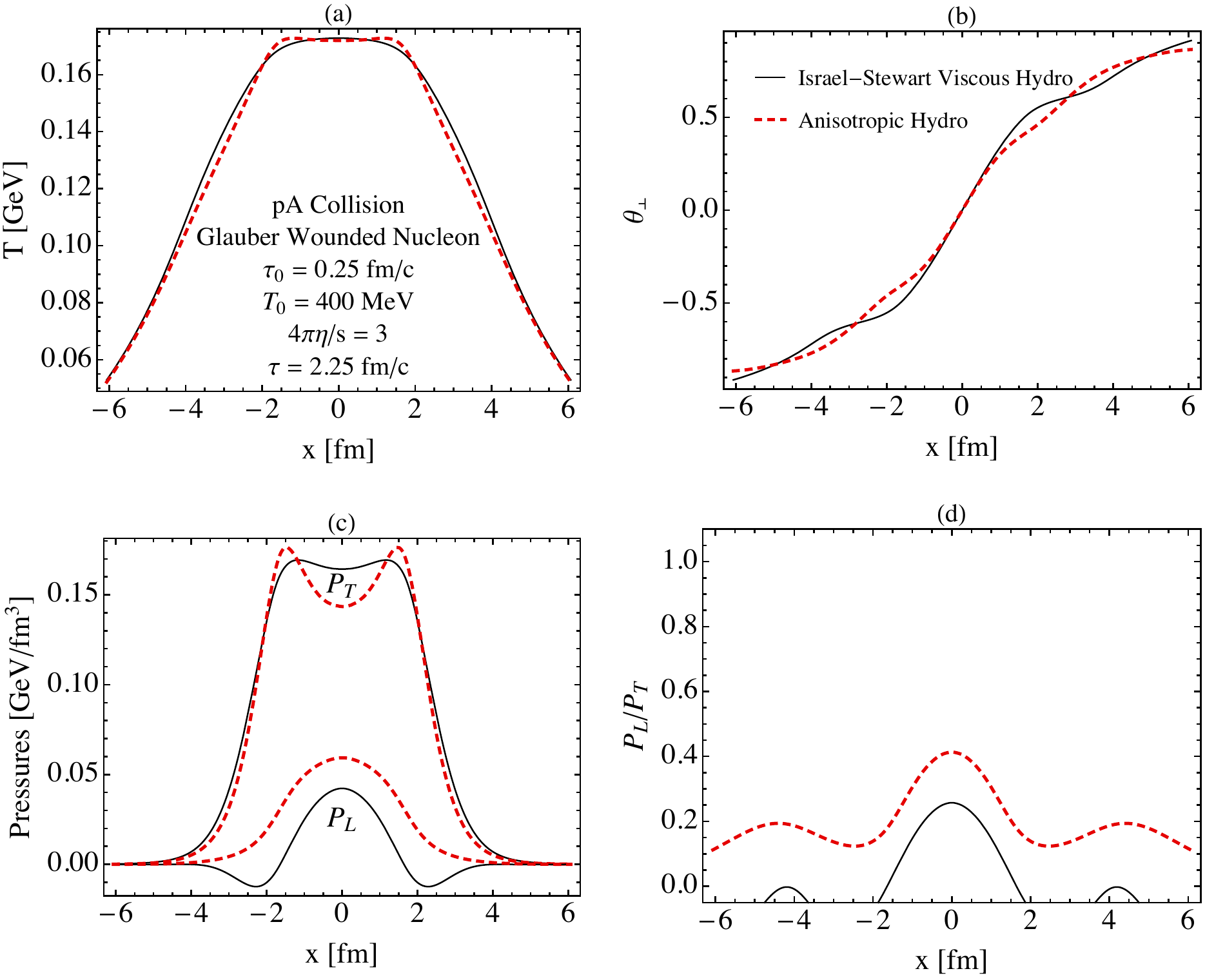}
\caption{Comparison of the numerical 1+1d solution to the Israel-Stewart and anisotropic hydrodynamics equations at $\tau = 2.25$ fm/c.  The initial central temperature was $T_0 = 400$ MeV at $\tau_0 = 0.25$ fm/c with $4\pi\eta/s = 3$.  The quantities shown in the panels are the same as in Fig.~\ref{fig:vacomp1}.}
\label{fig:vacomp3-pA}
\end{figure}

\begin{figure}[t!]
\includegraphics[width=1\linewidth]{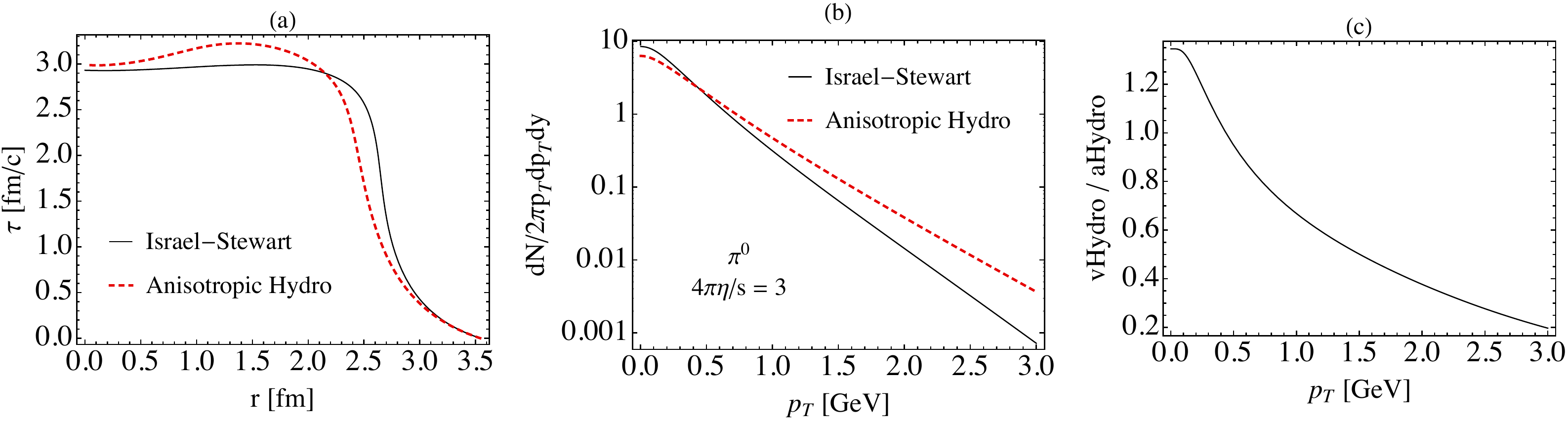}
\caption{Comparison of the numerical 1+1d solution to the Israel-Stewart and anisotropic hydrodynamics equations.  The parameters are the same as in Fig.~\ref{fig:vacomp3-pA}.  The quantities shown in the panels are the same as in Fig.~\ref{fig:vacompfreeze1}.}
\label{fig:vacompfreeze3-pA}
\end{figure}

\begin{figure}[t!]
\includegraphics[width=0.97\linewidth]{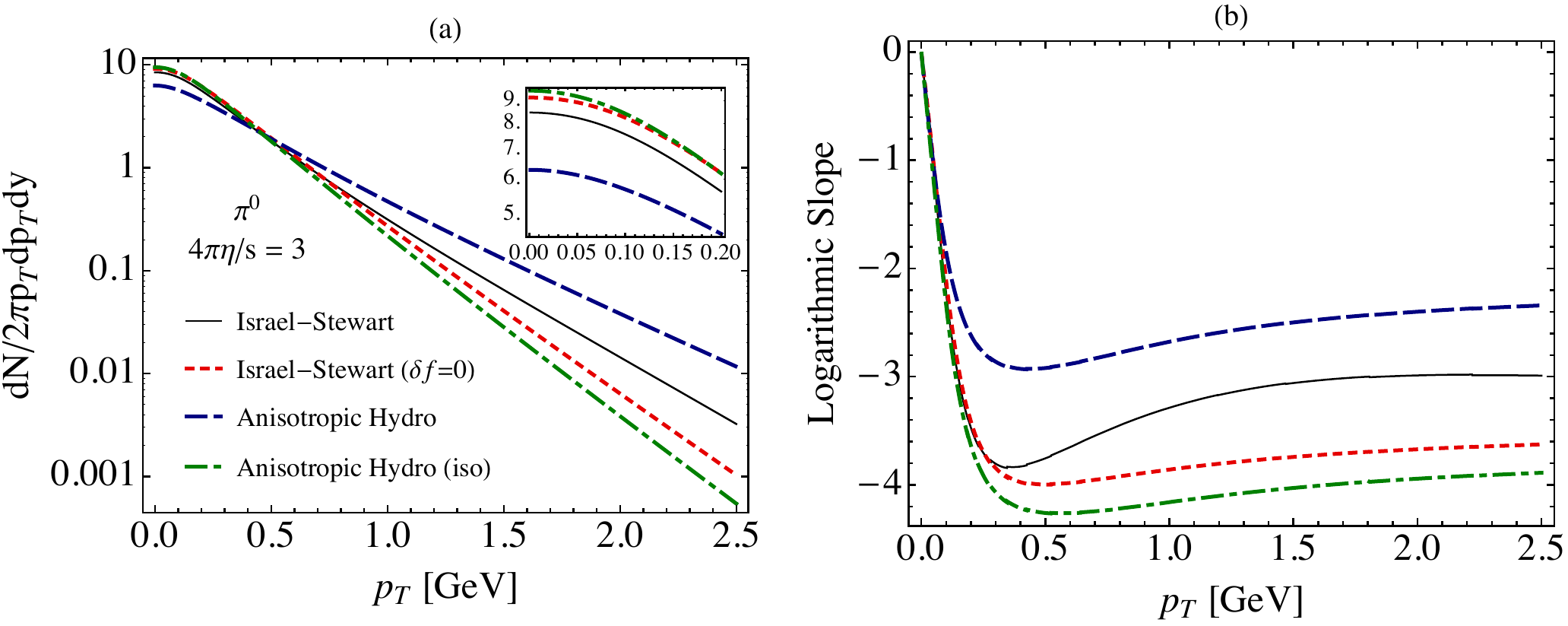}
\caption{Comparison of the p-Pb $\pi^0$ spectra resulting from Israel-Stewart hydrodynamics with and without ($\delta f=0$) viscous corrections to the distribution function, and anisotropic hydrodynamics with a complete treatment and assuming isotropic freeze-out using the effective temperature.  In this figure, we use the same conditions and parameters as shown in Fig.~\ref{fig:vacomp3-pA}.  The two panels show (a) the resulting neutral pion spectra and (b) the logarithmic slope of the various curves.}
\label{fig:specComp2}
\end{figure}

Similar to the previous subsection, we use the Glauber wounded-nucleon overlap to set the initial energy density.  As our test case, we consider a p-Pb collisions.  We use the same numerical and physical parameters as in the case of Pb-Pb collisions, except here we consider a lower initial central temperature of $T_0 = 400$ MeV.
Our findings are shown in Figs.~\ref{fig:vacomp3-pA}-\ref{fig:specComp2}.  From Fig.~\ref{fig:vacomp3-pA} we see that, assuming $4\pi\eta/s = 3$, there are relatively small differences in the temperature and flow profiles at $\tau = 2.25$ fm/c.  However, we see quite significant differences in the transverse and longitudinal pressures, with second-order viscous hydrodynamics again predicting negative longitudinal pressure and a quite different longitudinal to transverse pressure ratio.  From Fig.~\ref{fig:vacompfreeze3-pA} we see that the two methods result in quite different particle spectra, even at low momentum.  Finally, in Fig.~\ref{fig:specComp2}, we present a comparison using different freeze-out prescriptions for both anisotropic hydrodynamics and second-order viscous hydrodynamics.  The panels and methods are the same as we previously described in the context of Fig.~\ref{fig:specComp1}.  As we can see from Fig.~\ref{fig:specComp2}, the inclusion of viscous (anisotropic) corrections to the distribution function has a significant effect for p-A collisions.  The two methods seem to agree qualitatively concerning the direction of the correction, but differ quantitatively.

Comparing the smaller p-Pb to the larger Pb-Pb collision system, we see that pressure anisotropy effects are larger in the smaller system, but that they are also more severely overestimated by the Israel-Stewart approach. By resumming the leading viscous effects into the leading-order distribution function, as is done in anisotropic hydrodynamics, the pressure anisotropies are significantly reduced and the validity of the hydrodynamic approach, especially in small collision systems, is significantly improved.  

\section{Conclusions}
\label{sec:conclusion}

In this paper we demonstrated how to (i) impose a realistic EoS and (ii) self-consistently perform hadronic freeze-out in the context of leading-order 1+1d anisotropic hydrodynamics. The methods used here to implement these ingredients can be straightforwardly extended to dynamical evolution in more than one spatial dimension. In the case of freeze-out, they can be also extended to account for non-ellipsoidal corrections to the LRF one-particle distribution function, based on the viscous anisotropic hydrodynamics formalism of Refs.~\cite{Bazow:2013ifa,Bazow:2015cha}.

We compared the results obtained with anisotropic hydrodynamics to results obtained with the widely-used Israel-Stewart framework.   In the limit of small $\eta/s$ we found that the two frameworks agree well as they should.
For Pb-Pb collisions with $4\pi\eta/s = 1$, the maximal differences in the pressure anisotropy and neutral pion spectra for $p_T < 3$ GeV were found to be approximately 2\%.  For larger values of $\eta/s$, we naturally saw larger differences.  For Pb-Pb collisions with $4\pi\eta/s = 10$, we found significant corrections to the pressure anisotropy and, for $p_T < 3$ GeV, we found a maximal effect of approximately 100\% for the neutral pion spectra.  In general, we found that, for Pb-Pb collisions, second-order viscous hydrodynamics predicts stronger radial flow, resulting in harder spectra and more dissipative particle production than anisotropic hydrodynamics.  

We also presented results for the total charged particle multiplicity $N_{\rm chg}$ as a function $\eta/s$.  We found that, while $N_{\rm chg}$ increases monotonically as a function of $\eta/s$ using second-order viscous hydrodynamics, anisotropic hydrodynamics predicts that there is maximum in $N_{\rm chg}$ around $4 \pi \eta/s \sim 9 - 11$ depending on the assumed initial temperature.  This observation is consistent with the expectation that dissipative particle production should vanish in the free-streaming limit.

Finally, we considered the case of p-A collisions.  In this case, we found quite significant differences between the two frameworks.  Since the anisotropic hydrodynamics equations used herein have been shown to better reproduce exact solutions to the Boltzmann equation in the relaxation time approximation, we have some reason to believe that the anisotropic hydrodynamics results obtained herein are more reliable than those obtained using Israel-Stewart second-order viscous hydrodynamics.  Our results show that viscous effects are smaller in anisotropic hydrodynamics than predicted by Israel-Stewart theory, which improves significantly the applicability of a hydrodynamic approach to such small collision systems.  We also demonstrated that the inclusion or exclusion of the viscous (anisotropic) corrections to the freeze-out one-particle distribution functions dramatically influences, e.g., the final neutral pion spectra.  This indicates that at freeze-out the system is not equilibrated and still quite anisotropic in the LRF.  We provided further evidence for this conclusion by computing the pressure anisotropy of the fluid for the case of p-A collisions, finding that both the anisotropic hydrodynamics and Israel-Stewart frameworks predict quite large pressure anisotropies even on the late-proper-time portion of the freeze-out hypersurface.  In this context, we also emphasize that using leading-order anisotropic hydrodynamics one is able to guarantee that the pressures and one-particle distribution functions are positive.

Looking to the future, we have demonstrated how to implement two critically needed components for the anisotropic hydrodynamics program.  The next steps will be to extend the codes to 2+1d and 3+1d evolution including also perturbatively the second-order corrections in the spirit of Refs.~\cite{Bazow:2013ifa,Bazow:2015cha}.  Of course, one could already use the codes developed for use in this paper to attempt phenomenological fits of data coming from central Pb-Pb and p-Pb collisions at RHIC and LHC.  For this, one merely needs to add a hadronic afterburner and perform some fitting.  We postpone this phenomenological exercise to a future publication.

\acknowledgments{ 
We thank M. Al Qahtani, G. Denicol, and B. Schenke for useful conversations.  M. Strickland and M. Nopoush were supported in part by the U.S. Department of Energy, Office of Science, Office of Nuclear Physics under Awards No.~DE-SC0013470 and (within the framework of the JET Collaboration) No.~DE-AC0205CH11231.  In addition, M. Strickland thanks the Kavli Institute for Theoretical Physics China and the Chinese Academy of Sciences under the auspices of the ``sQGP and extreme QCD'' program.  R. Ryblewski was supported by the Polish National Science Center Grant No.~DEC-2012/07/D/ST2/02125.  D. Bazow, U. Heinz, and M. Martinez were supported by the U.S. Department of Energy, Office of Science, Office of Nuclear Physics under Awards No.~DE-SC0004286 and (within the framework of the JET Collaboration) No.~DE-SC0004104.    
}

\appendix

\section{Minkowski basis vectors}
\label{app:basis}

For azimuthally-symmetric and longitudinally boost-invariant flow, one can parameterize the orthogonal basis vectors in lab frame cartesian Minkowski coordinates as
\be
\begin{array}{ll}
\begin{aligned}
u^0 &= \cosh\theta_\perp \cosh\varsigma \, , \\
u^1 &= \sinh\theta_\perp \cos\phi \, , \\
u^2 &= \sinh\theta_\perp \sin\phi \, , \\
u^3 &= \cosh\theta_\perp \sinh\varsigma \, ,
\end{aligned}
\quad \quad \quad &
\begin{aligned}
X^0 &= \sinh\theta_\perp \cosh\varsigma \, , \\
X^1 &= \cosh\theta_\perp \cos\phi \, , \\
X^2 &= \cosh\theta_\perp \sin\phi \, , \\
X^3 &= \sinh\theta_\perp \sinh\varsigma \, ,
\end{aligned}
\\
&\vspace{-3mm}\\
\begin{aligned}
Y^0 &= 0 \, , \\
Y^1 &= -\sin\phi \, , \\
Y^2 &= \cos\phi \, , \\
Y^3 &= 0 \, ,
\end{aligned}
\quad \quad  \quad &
\begin{aligned}
Z^0 &= \sinh\varsigma \, , \\
Z^1 &= 0 \, , \\
Z^2 &= 0 \, , \\
Z^3 &= \cosh\varsigma \, .
\\
\end{aligned}
\end{array}
\label{eq:expvectorbasis}
\ee

In the local rest frame (LRF) they are simply
\ba
u_{\rm LRF}^\mu &=&(1,0,0,0) \, ,\nonumber\\
X_{\rm LRF}^\mu &=&(0,1,0,0) \, ,\nonumber\\
Y_{\rm LRF}^\mu &=&(0,0,1,0) \, ,\nonumber\\
Z_{\rm LRF}^\mu &=&(0,0,0,1) \, .
\ea

\section{Formulas for derivatives}
\label{app:derivatives}

In this section, the identities for derivatives and divergences for 1+1d boost-invariant and azimuthally-symmetric flow are summarized.

\vspace{3mm}

\paragraph*{\bf Directional derivatives}
\ba 
D_u &\equiv & u \cdot \partial = \cosh\theta_\perp \partial_\tau + \sinh\theta_\perp \partial_r \, ,\nonumber\\
D_x &\equiv& X \cdot \partial = \sinh\theta_\perp \partial_\tau + \cosh\theta_\perp \partial_r \, , \nonumber\\
D_y &\equiv& Y \cdot \partial = \frac{1}{r} \partial_\phi \, ,\nonumber \\
D_z &\equiv& Z \cdot \partial = \frac{1}{\tau} \partial_{\varsigma} \, .
\label{useful-eqns-1}
\ea
\paragraph*{\bf Divergences}
\ba 
\theta_u &\equiv& \partial \cdot u =
\cosh\theta_\perp \left(\frac{1}{\tau} + \partial_r \theta_\perp \right)+ \sinh\theta_\perp \left( \frac{1}{r} + \partial_\tau \theta_\perp \right),  \nonumber \\
\theta_x &\equiv& \partial \cdot X = \sinh\theta_\perp \left(\frac{1}{\tau} + \partial_r \theta_\perp \right)
+ \cosh\theta_\perp \left( \frac{1}{r} + \partial_\tau \theta_\perp \right), \nonumber \\
\theta_y &\equiv& \partial \cdot Y = 0 \, , \nonumber \\
\theta_z &\equiv& \partial \cdot Z = 0 \, .
\label{useful-eqns-2}
\ea
\paragraph*{\bf Convective derivatives}
\ba 
D_u u  &\equiv & (u \cdot \partial) u = X \left(\cosh\theta_\perp \partial_\tau \theta_\perp + \sinh\theta_\perp \partial_r \theta_\perp \right), \nonumber \\
D_u X &\equiv & (u \cdot \partial) X =
u \left(\cosh\theta_\perp \partial_\tau \theta_\perp + \sinh\theta_\perp \partial_r \theta_\perp \right),  \nonumber \\
D_u Y &\equiv & (u \cdot \partial) Y = 0 \, , \nonumber \\
D_u Z&\equiv & (u \cdot \partial) Z = 0 \, .
\label{useful-eqns-3}
\ea
\paragraph*{\bf Directional derivatives}
\ba 
D_x u &\equiv&(X \cdot \partial) u =
X \left(\sinh\theta_\perp \partial_\tau \theta_\perp + \cosh\theta_\perp \partial_r \theta_\perp \right), \nonumber \\  
D_x X &\equiv&(X \cdot \partial) X = u \left(\sinh\theta_\perp \partial_\tau \theta_\perp + \cosh\theta_\perp \partial_r \theta_\perp \right),  \nonumber \\
D_x Y &\equiv& (X \cdot \partial) Y = 0 \, , \nonumber \\
D_x Z &\equiv& (X \cdot \partial) Z = 0 \, ,
\label{useful-eqns-4}
\ea
\ba 
D_y u &\equiv& (Y \cdot \partial) u = \frac{\sinh\theta_\perp}{r} \, Y \, ,  \nonumber \\
D_y X &\equiv&(Y \cdot \partial) X = \frac{\cosh\theta_\perp}{r} \, Y  \, , \nonumber \\
D_y Y &\equiv&(Y \cdot \partial) Y = \frac{1}{r}\left(u\sinh\theta_\perp - X\cosh\theta_\perp \right) ,  \nonumber \\
D_y Z &\equiv&(Y \cdot \partial) Z =0 \, ,
\label{useful-eqns-5}
\ea
\ba 
D_z u &\equiv & (Z \cdot \partial) u = \frac{\cosh\theta_\perp}{\tau} \, Z \, ,  \nonumber \\
D_z X &\equiv & (Z \cdot \partial) X = \frac{ \sinh\theta_\perp}{\tau} \, Z  \, ,  \nonumber \\
D_z Y &\equiv & (Z \cdot \partial) Y = 0 \, , \nonumber \\
D_z Z &\equiv & (Z \cdot \partial) Z = \frac{1}{\tau}\left(u\cosh\theta_\perp - X\sinh\theta_\perp \right).
\label{useful-eqns-6}
\ea

\section{Special Functions}
\label{app:special-functions}

In this appendix we list the ${\cal R}$ and ${\cal H}$ functions appearing in Sec.~\ref{sec:equations}.  They are \cite{Tinti:2013vba,Nopoush:2014qba}
\ba
{\cal R}(\alpha_x,\alpha_z)&=&\frac{\alpha_x\alpha_y}{4\pi}\int_0^{2\pi}d\phi\,\alpha_\perp^2 {\cal H}_2\left(\frac{\alpha_z}{\alpha_\perp}\right) , \label{eq:rfunc} \\
{\cal H}_{Tx}(\alpha_x,\alpha_z)&=&\frac{3\alpha_x\alpha_y}{4\pi}\int_0^{2\pi}d\phi\,\alpha_x^2 \cos^2\phi\, {\cal H}_{2T}\left(\frac{\alpha_z}{\alpha_\perp}\right) , \label{eq:htxfunc} \\
{\cal H}_{Ty}(\alpha_x,\alpha_z)&=&\frac{3\alpha_x\alpha_y}{4\pi}\int_0^{2\pi}d\phi\,\alpha_y^2 \sin^2\phi\, {\cal H}_{2T}\left(\frac{\alpha_z}{\alpha_\perp}\right) , \label{eq:htyfunc} \\
{\cal H}_L(\alpha_x,\alpha_z)&=&\frac{3\alpha_x\alpha_y}{4\pi}\int_0^{2\pi}d\phi\,\alpha_\perp^2 {\cal H}_{2L}\left(\frac{\alpha_z}{\alpha_\perp}\right) ,
\label{eq:hlfunc}
\ea 
with $\alpha_\perp \equiv \sqrt{\alpha_x^2 \cos^2\phi + \alpha_y^2 \sin^2\phi}$ and
\ba
{\cal H}_2(y) &=&  \frac{y}{\sqrt{y^2-1}}\left(\tanh^{-1}\!\frac{\sqrt{y^2-1}}{y}+y\sqrt{y^2-1}\right) ,
\label{eq:H2}
\\
{\cal H}_{2T}(y) &=& \frac{y}{(y^2-1)^{3/2}}\left((2y^2-1)\tanh^{-1}\!\frac{\sqrt{y^2-1}}{y}-y\sqrt{y^2-1}\right) , 
\label{eq:H2T}
\\
{\cal H}_{2L}(y) &=& \frac{y^3}{(y^2-1)^{3/2}}\left(y\sqrt{y^2-1}-\tanh^{-1}\!\frac{\sqrt{y^2-1}}{y}\right) .
\label{eq:H2L}
\ea

\section{Second-order viscous hydrodynamics}
\label{app:vhydro}

As with anisotropic hydrodynamics, the viscous hydrodynamics dynamical equations can be obtained by taking moments of the Boltzmann equation. Taking the first and second moments of Boltzmann equation one obtains \cite{Baier:2007ix,Denicol:2012cn}
\ba
(\varepsilon+P)D_u u^\mu&=&\nabla^\mu P-\Delta^\mu_{\nu} \nabla_\sigma
\pi^{\nu \sigma}+\pi^{\mu \nu} D_u u_\nu\, ,
\label{eq:vhyd1-1}\\
D_u \varepsilon &=& - (\varepsilon+P)\nabla_\mu u^\mu + \pi^{\mu \nu}
\sigma_{\mu\nu}\, ,
\label{eq:vhyd2-1}\\
\tau_{\pi} (\Delta^{\mu\nu}_{\alpha\beta} D_u \pi^{\alpha \beta}+\frac{4}{3}\pi^{\mu\nu}\theta_u)
+\pi^{\mu \nu}&=&  2 \eta  \sigma^{\mu\nu}
- \tau_{\pi\pi}\pi^{\alpha\langle\mu}\sigma^{\nu\rangle}_{\ \ \alpha}
+ 2 \tau_{\pi} \pi^{\alpha \langle\mu}\omega^{\nu\rangle}_{\ \ \alpha} \, ,
\label{eq:vhyd3-1}
\ea
where $\varepsilon\equiv\varepsilon_{\rm eq}$ and $P\equiv P_{\rm eq}$ are the equilibrium (isotropic) energy density and pressure, respectively, $\tau_\pi$ is the shear relaxation time, and $\tau_{\pi\pi}$ is the shear-shear-coupling transport coefficient.  The various notations used are 
\ba
d_\mu u^\nu &\equiv& \partial_\mu u^\nu+\Gamma_{\mu\alpha}^{\nu} u^\alpha \, , \nonumber \\
D_u &\equiv& u_\mu d^\mu \, , \nonumber \\
\nabla^\mu &\equiv& \Delta^{\mu \nu} d_\nu \, , \nonumber \\ 
\omega^{\mu \nu} &\equiv& \frac{1}{2} ( \nabla^\mu u^\nu - \nabla^\nu u^\mu ) \, ,
\label{eq:identities1}
\ea
The non-vanishing Christoffel symbols for polar Milne coordinates are $\Gamma^\tau_{\varsigma\varsigma}=\tau$, $\Gamma^\varsigma_{\varsigma\tau}=1/\tau$, $\Gamma^r_{\phi\phi}=-r$, and $\Gamma^\phi_{r\phi}=1/r$.
For the smooth initial conditions considered herein, the vorticity is zero at all times.  We also set the transport coefficient $\tau_{\pi\pi}$ to zero since this has been done in almost all other implementations to date (see, however, \cite{Ryu:2015vwa}). As a result, we drop the last two terms in Eq.~(\ref{eq:vhyd3-1}).

\subsection*{1+1d equations of motion}

In the boost-invariant and cylindrically-symmetric case, one has $u^\mu = (u^\tau,u^r,0,0)$ and, as a result, $v \equiv \tanh\theta_\perp = u^r/u^\tau$.  In addition, for this case, the shear tensor has the following form
\be
\pi^{\mu\nu}=
\begin{pmatrix}
\pi^{\tau\tau}&\pi^{\tau r}&0&0\\ \pi^{\tau r}&\pi^{rr}&0&0\\0&0&\pi^{\phi\phi}&0\\0&0&0&\pi^{\varsigma\varsigma}
\end{pmatrix} .
\label{eq:pi-mat-1}
\ee
Note also that, in the boost-invariant and cylindrically-symmetric case, $D_u$ as defined in Eq.~(\ref{eq:identities1}) reduces to the expression presented previously in Eq.~(\ref{useful-eqns-1}).

In this case, expanding Eqs.~(\ref{eq:vhyd1-1}), (\ref{eq:vhyd2-1}), and (\ref{eq:vhyd3-1}) in polar Milne coordinates one obtains five independent equations
\ba
(\varepsilon+P)D_u u^\tau &=&-(u^r)^2\left(\partial_\tau P
-d_\nu \pi^\nu_\tau\right) 
-u^\tau u^r \left(\partial_r P-d_\nu \pi^\nu_r\right) , \label{eq:ut}\\
(\varepsilon+P)D_u u^r &=& -u^\tau u^r\left(\partial_\tau P-d_\nu \pi^\nu_\tau\right)
-(u^\tau)^2 \left(\partial_r P-d_\nu\pi^\nu_r\right) , \label{eq:ur} \\
D_u\varepsilon &=& -(\varepsilon+P) \theta_u 
-\pi^r_r (1-v^2)^2\nabla^{\langle r} u^{r\rangle}-r^2\, \pi^\phi_\phi \nabla^{\langle\phi} u^{\phi\rangle}
-\tau^2\,\pi^\varsigma_\varsigma \nabla^{\langle\varsigma} u^{\varsigma \rangle} , \hspace{1cm} \label{eq:e}
\ea


\ba
\tau_{\pi} (D_u \pi^\phi_\phi+\frac{4}{3}\theta_u\pi^\phi_\phi) &=& 
-2\eta\, r^2\,  \nabla^{\langle\phi}
u^{\phi \rangle} - \pi^\phi_\phi  \, ,  \label{eq:pi-ff}  \\
\tau_{\pi} (D_u \pi^\varsigma_\varsigma+\frac{4}{3}\theta_u\pi^\varsigma_\varsigma) &=& 
-2\eta\, \tau^2\,  \nabla^{\langle\varsigma}
u^{\varsigma \rangle} - \pi^\varsigma_\varsigma  \, ,\label{eq:pi-vv}
\ea
where 
\ba
- d_\nu \pi^\nu_\tau &=& v^2 \partial_\tau \pi^r_r+v \partial_r \pi^r_r+\pi^r_r\left(\partial_\tau v^2+\partial_r v+\frac{v^2}{\tau}
+\frac{v}{r}\right)+\frac{1}{\tau}\pi^\varsigma_\varsigma \, , \label{eq:dpi-tau}\\
d_\nu \pi^\nu_r &=& v\, \partial_\tau \pi^r_r+\partial_r \pi^r_r+\pi^r_r \left(\partial_\tau v+\frac{v}{\tau}+\frac{2-v^2}{r}\right)
+\frac{1}{r} \pi^\varsigma_\varsigma \, . \label{eq:dpi-r}
\ea
In addition, one needs the following
\ba 
\nabla^{\langle r} u^{r \rangle} &=& - \partial_r u^r- u^r D_u u^r +\frac{1}{3} (u^\tau)^2 \theta_u \, ,\label{eq:nablar-ur}\\ 
r^2 \nabla^{\langle \phi} u^{\phi \rangle} &=&- \frac{u^r}{r}+\frac{1}{3}\theta_u \, , \label{eq:nablavph-uph} \\  
\tau^2 \nabla^{\langle \varsigma} u^{\varsigma \rangle} &=&- \frac{u^\tau}{\tau}+\frac{1}{3}\theta_u \, , \label{eq:nablav-uv} \\
\theta_u\equiv\nabla_\alpha u^\alpha &=&d_\alpha u^\alpha = \partial_\tau u^\tau+\partial_r u^r+\frac{u^\tau}{\tau}+\frac{u^r}{r}\, ,
\ea
where $\pi^r_\tau=-v\, \pi^r_r$ and $\pi^\phi_\phi=-\pi^\varsigma_\varsigma-(1-v^2) \pi^r_r$ which are a consequence of the transversality of the shear-stress tensor, $u_\mu \pi^{\mu\nu}=0$.  This system of equations has to be closed by providing an equation of state (EoS), e.g. $P_{\rm eq}=P_{\rm eq}(\varepsilon_{\rm eq})$.  For the numerical results presented in the body of the manuscript, we use the lattice-based EoS specified in Sec.~\ref{sec:eos}.

\subsection*{Viscous hydrodynamics freeze-out}

The distribution function on the freeze-out hypersurface can be computed assuming that there is a linear correction to the equilibrium distribution function \cite{Teaney:2003kp}
\be
f(p,x)=f_{\rm eq} \Bigg[1+(1 - a f_{\rm eq}) \frac{p_\mu p_\nu \pi^{\mu\nu}}{2(\varepsilon+P)T^2}\Bigg] .
\label{eq:vfreeze}
\ee
Using tensor transformations applied to Eq.~(\ref{eq:ptl-mom}) the components of the four-momentum in polar Milne coordinates are
\ba
p^\tau &=&p^t\cosh\varsigma-p^z\sinh\varsigma=m_\perp\cosh(y-\varsigma) \, , \nonumber \\
p^r&=&p^x \cos\phi+p^y\sin\phi=p_\perp\cos(\phi-\varphi) \, , \nonumber \\
p^\phi&=&-p^x\frac{\sin\phi}{r}+p^y\frac{\cos\phi}{r}=-\frac{p_\perp}{r}\sin(\phi-\varphi)\, , \nonumber \\
p^\varsigma&=&-p^t\frac{\sinh\varsigma}{\tau}+p^z\frac{\cosh\varsigma}{\tau}=\frac{m_\perp}{\tau}\sinh(y-\varsigma) \, .
\label{eq:p-con}
\ea

Using Eq.~(\ref{eq:pi-mat-1}) and expanding $p_\mu p_\nu\pi^{\mu\nu}$ in polar Milne coordinates one has
\ba 
p_\mu p_\nu \pi^{\mu\nu}&=&p_\tau^2\pi^{\tau\tau}+p_\tau p_r\pi^{\tau r}+p_r p_\tau\pi^{r\tau}+p_r^2\pi^{rr}+p_\phi^2\pi^{\phi\phi}+p_\varsigma^2\pi^{\varsigma\varsigma} \, .
\label{eq:expansion1}
\ea
As a result, one obtains
\ba 
p_\mu p_\nu \pi^{\mu\nu}=&-&\left(\frac{\pi^\phi_\phi+\pi^\varsigma_\varsigma}{v^2-1}\right)\Big(m_\perp v \cosh(y-\varsigma)-p_\perp\cos(\phi-\varphi)\Big)^2 \nonumber \\
&-&\pi^\phi_\phi p^2_\perp\sin^2(\phi-\varphi)-\pi^\varsigma_\varsigma m^2_\perp\sinh^2(y-\varsigma) \, .
\ea

\section{Factorization in non-conformal anisotropic hydrodynamics}
\label{app:factorization}

In the case that the particles comprising the system are massive, conformality is broken and it is no longer possible to multiplicatively factorize the energy density and pressures as in Eqs.~(\ref{eq:ed})-(\ref{eq:pz}) \cite{Martinez:2012tu,Florkowski:2014sfa,Florkowski:2014bba,Nopoush:2014pfa}.  For a system with constant mass $m$ one has instead~\cite{Nopoush:2014pfa}
\ba
\varepsilon &=& {\cal H}_3({\boldsymbol\alpha},\Phi,\hat{m}) \, \lambda^4 \, , \\
P_T &=& {\cal H}_{3T}({\boldsymbol\alpha},\Phi,\hat{m}) \, \lambda^4 \, , \\
P_L &=& {\cal H}_{3L}({\boldsymbol\alpha},\Phi,\hat{m}) \, \lambda^4 \, ,
\label{eq:massivet}
\ea
where ${\boldsymbol\alpha} = (\alpha_x,\alpha_y,\alpha_z)$, $\hat{m} \equiv m/\lambda$, $\hat{p} \equiv p/\lambda$, and the constraint $\Phi = \frac{1}{3} \sum_i \alpha_i^{-2} - 1$ with $i \in \{x,y,z\}$ is implicit.  The functions ${\cal H}_3$, ${\cal H}_{3T}$, and ${\cal H}_{3L}$ are given by
\ba
{\cal H}_3({\boldsymbol\alpha},\Phi,\hat{m}) &\equiv& \tilde{N} \alpha_x \alpha_y
\int_0^{2\pi} d\phi \, \alpha_\perp^2 \int_0^\infty d\hat{p} \, \hat{p}^3  f_{\rm eq}\!\left(\!\sqrt{\hat{p}^2 + \hat{m}^2}\right) {\cal H}_2\!\left(\frac{\alpha_z}{\alpha_\perp},\frac{\hat{m}}{\alpha_\perp \hat{p}} \right) ,
\label{eq:h3gen} \\
{\cal H}_{3T}({\boldsymbol\alpha},\Phi,\hat{m}) &\equiv&  \frac{1}{2} \tilde{N}\alpha_x \alpha_y
\int_0^{2\pi} d\phi \,\alpha_\perp^2 \int_0^\infty d\hat{p} \, \hat{p}^3  f_{\rm eq}\!\left(\!\sqrt{\hat{p}^2 + \hat{m}^2}\right) {\cal H}_{2T}\!\left(\frac{\alpha_z}{\alpha_\perp},\frac{\hat{m}}{\alpha_\perp \hat{p}} \right) ,
\label{eq:h3tgen} \\
{\cal H}_{3L}({\boldsymbol\alpha},\Phi,\hat{m}) &\equiv&  \tilde{N} \alpha_x \alpha_y
\int_0^{2\pi} d\phi \, \alpha_\perp^2 \int_0^\infty d\hat{p} \, \hat{p}^3  f_{\rm eq}\!\left(\!\sqrt{\hat{p}^2 + \hat{m}^2}\right) {\cal H}_{2L}\!\left(\frac{\alpha_z}{\alpha_\perp},\frac{\hat{m}}{\alpha_\perp \hat{p}} \right) ,
\label{eq:h3lgen}
\ea
with $\alpha_\perp^2 \equiv \alpha_x^2 \cos^2\phi + \alpha_y^2 \sin^2\phi$, $\hat{p} = |\hat{\bf p}|$, $\tilde{N} = N_{\rm dof}/(2\pi)^3$ with $N_{\rm dof}$ being the number of degrees of freedom, and the functions ${\cal H}_2$, ${\cal H}_{2T}$, and ${\cal H}_{2L}$ are given by
\ba
 {\cal H}_2(y,z) &=& \frac{y}{\sqrt{y^2-1}} \left( (1+z^2)
\tanh^{-1} \sqrt{\frac{y^2-1}{y^2+z^2}} + \sqrt{(y^2+z^2)(y^2-1)} \, \right) ,
\label{eq:H2m} \\
{\cal H}_{2T}(y,z) &=& \frac{y}{(y^2-1)^{3/2}}
\left[\left(z^2+2y^2-1\right) 
\tanh^{-1}\sqrt{\frac{y^2-1}{y^2+z^2}}
-\sqrt{(y^2-1)(y^2+z^2)} \right] , 
\label{eq:H2Tm} \\
{\cal H}_{2L}(y,z) &=& \frac{y^3}{(y^2-1)^{3/2}}
\left[
\sqrt{(y^2-1)(y^2+z^2)}-(z^2+1)
\tanh^{-1}\sqrt{\frac{y^2-1}{y^2+z^2}} \,\,\right] . 
\label{eq:H2Lm}
\ea
%

\begin{figure}[t]
\includegraphics[width=0.32\linewidth]{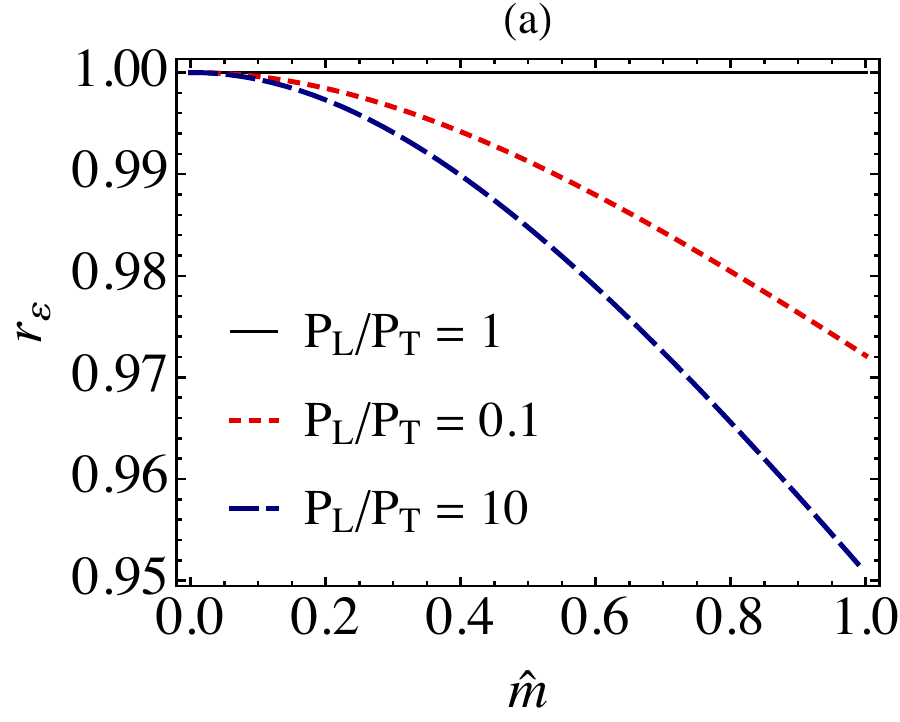}
\includegraphics[width=0.33\linewidth]{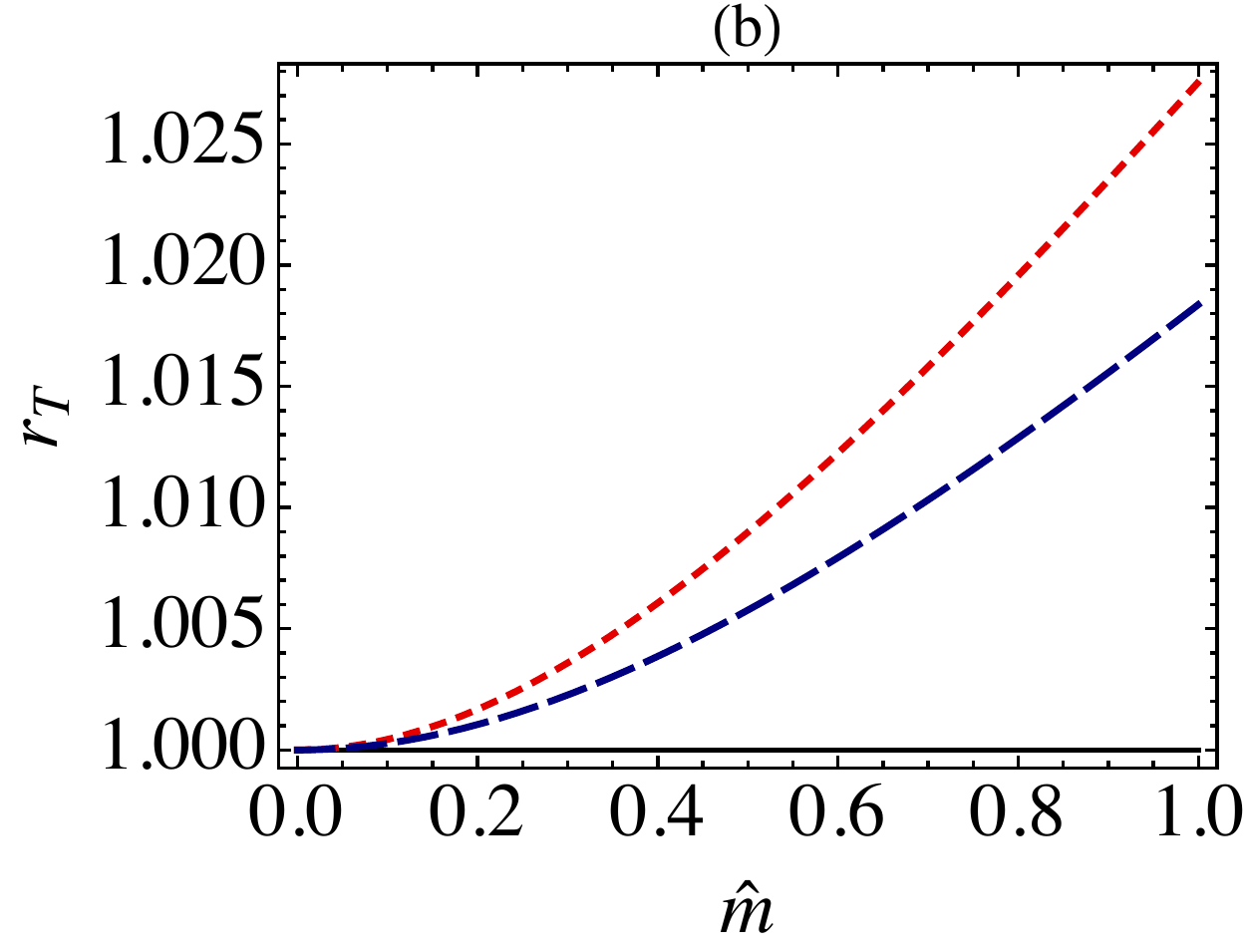}
\includegraphics[width=0.32\linewidth]{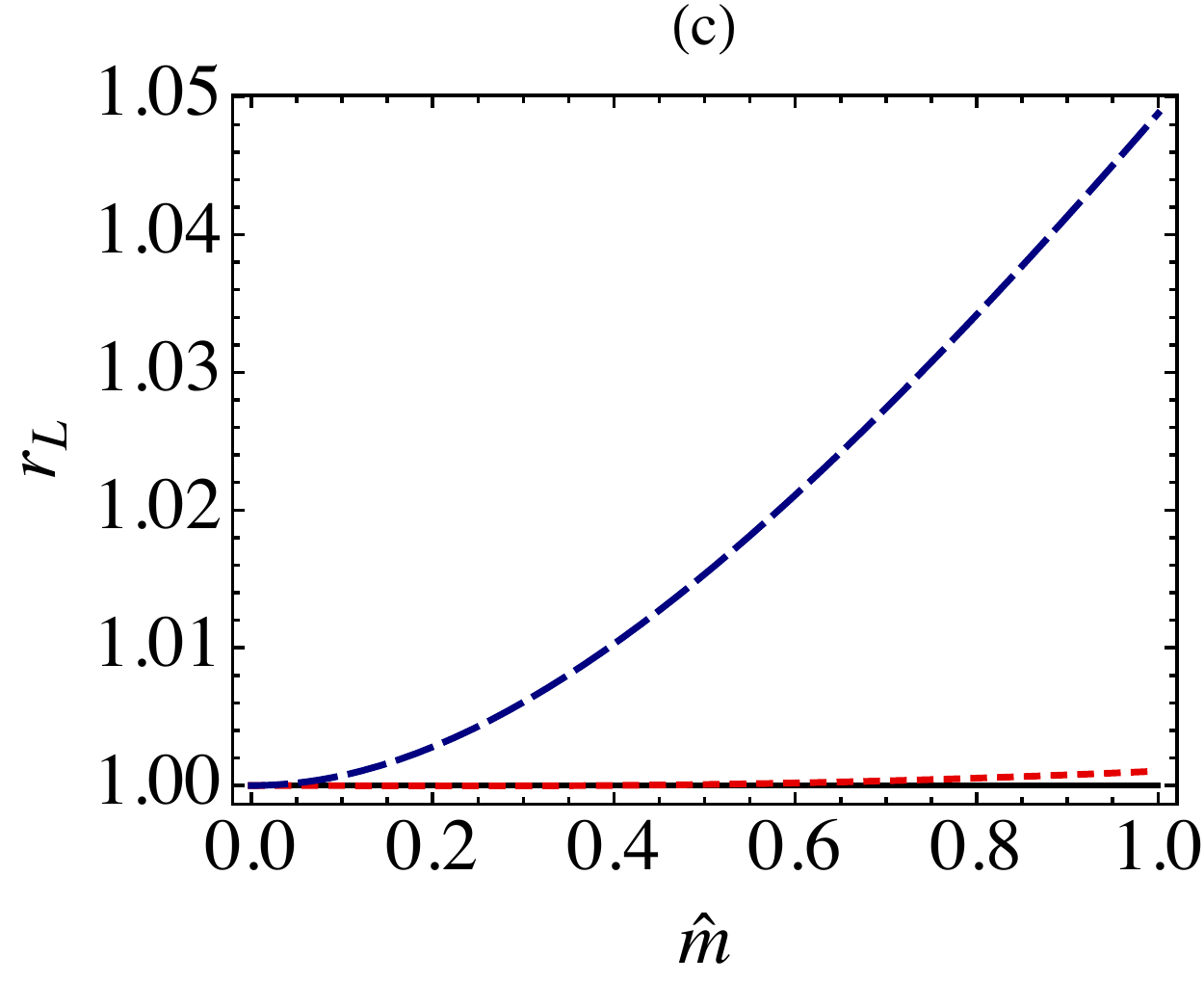}
\caption{The ratios (a) $r_\varepsilon$, (b) $r_T$, and (c) $r_L$ as a function of $\hat{m} = m/\lambda$ for three different values of the pressure anisotropy $P_L/P_T = \{0.1, 1, 10\}$ corresponding to $\alpha_z = \{0.65499, 1,5.4631\}$.}
\label{fig:rats}
\end{figure}

Since ${\cal H}_3$, ${\cal H}_{3T}$, and ${\cal H}_{3L}$ depend on many variables, we first restrict ourselves to the case that $\alpha_x = \alpha_y$ which is appropriate for the case of 1+1d dynamics considered in the body of the text.  In addition, since, in the body of the text we assume $\Phi=0$, we also assume this here.  With these assumptions, one can write $\alpha_x$ in terms of $\alpha_z$~\cite{Nopoush:2014pfa}
\be
\alpha_x^2 = \frac{2 \alpha_z^2}{3 \alpha_z^2 - 1} \, .
\ee
In order to test the degree to which the conformal factorization occurs in the massive case, we define three ratios
\ba
r_\varepsilon &\equiv& \frac{{\cal H}_3(\boldsymbol\alpha,0,\hat{m})}{{\cal H}_3({\bf 1},0,\hat{m}) {\cal R}(\alpha_x,\alpha_z)} \, , \\
r_T &\equiv& \frac{{\cal H}_{3T}(\boldsymbol\alpha,0,\hat{m})}{{\cal H}_{3T}({\bf 1},0,\hat{m}) {\cal H}_{Tx}(\alpha_x,\alpha_z)} \, , \\
r_L &\equiv& \frac{{\cal H}_{3L}(\boldsymbol\alpha,0,\hat{m})}{{\cal H}_{3L}({\bf 1},0,\hat{m}) {\cal H}_{L}(\alpha_x,\alpha_z)} \, ,
\ea
where ${\cal R}$, ${\cal H}_{Tx}$, and ${\cal H}_L$ are defined in Eqs.~(\ref{eq:rfunc})-(\ref{eq:hlfunc}).  The factors of ${\cal H}_3({\bf 1},0,\hat{m})$, ${\cal H}_{3T}({\bf 1},0,\hat{m})$, ${\cal H}_L({\bf 1},0,\hat{m})$ are introduced in the denominator in order to compensate for the trivial mass dependence of the EoS in the isotropic case.  If these ratios are equal to one, then there is perfect factorization of the diagonal components of the energy momentum tensor and the size of the deviation from unity is indicative of the degree to which factorization is broken in the non-conformal case.  We present our numerical evaluation of these three ratios in Fig.~\ref{fig:rats}.  As can be seen from the three panels, as long as $0.1 < P_L/P_T < 10$ and $\hat{m} < 1$, the maximum correction to all of these ratios is approximately 5\%.  For the case of oblate pressure anisotropies, which is more relevant for phenomenological application to heavy ion collisions, the maximum correction is approximately 3\%.

\section{Gubser flow tests}
\label{app:tests}

In this appendix we present results of code tests for both the anisotropic hydrodynamics and viscous hydrodynamics codes used in the body of the text.  For our tests, we initialize the codes at a given proper time using the exact Gubser solution of the hydrodynamic equations appropriate for each case (see \cite{Nopoush:2014qba,Marrochio:2013wla}).\footnote{These exact solutions are the ones appropriate for each approximation scheme and are not to be confused with e.g. the exact solution of the Boltzmann equation in the relaxation time approximation (RTA) \cite{Denicol:2014xca,Denicol:2014tha}.  When compared to the exact RTA solution, anisotropic hydrodynamics does a dramatically better job reproducing the exact RTA solution and can be shown to analytically reproduce both the ideal and free-streaming limits \cite{Nopoush:2014qba}.}  Gubser flow is a conformal flow which, in polar Milne coordinates, is completely determined by symmetry constraints to be \cite{Gubser:2010ui,Gubser:2010ze}
\be 
u^\mu = (\cosh\theta_\perp,\sinh\theta_\perp,0,0) \, ,
\label{eq:gubser-flow}
\ee
where
\be 
\theta_\perp=\tanh^{-1} \left(\frac{2q^2\tau r}{1+q^2\tau^2+q^2r^2}\right) ,
\label{eq:thetaperp}
\ee
with $q$ being an energy scale which sets the transverse spatial size of the system.
Gubser flow is best understood through the introduction of de Sitter variables $\rho$ and $\theta$ \cite{Gubser:2010ui}
\ba 
\sinh{\rho} &=&  - \frac{1-q^2{\tau^2}+q^2r^2}{2q{\tau}}\, ,
\label{eq:desitter1} \\
\tan{\theta} &=& \frac{2qr}{1+q^2{\tau}^2-q^2r^2}\, ,
\label{eq:desitter2}
\ea
where $\rho$ and $\theta$ are components of de Sitter coordinates, $\hat{x}^\mu=(\rho,\,\theta,\,\phi,\,\varsigma)$.  Note that, for fixed $r$, the limit $\tau \rightarrow 0^+$ corresponds to the limit $\rho \rightarrow -\infty$ and the limit $\tau\rightarrow\infty$ corresponds to the limit $\rho \rightarrow \infty$.  As a consequence, the de Sitter map $-\infty<\rho<+\infty$ covers the future (forward) light cone.  In what follows in this appendix, all Weyl-rescaled variables defined in de Sitter coordinates are indicated with a hat.  Note that, since Gubser flow is conformal, the EoS used in the tests presented below is a conformal (ideal) equation state.

\subsection*{Gubser flow using Israel-Stewart second-order viscous hydrodynamics}
\label{app:gubservhydro}

\begin{figure}[t]
\includegraphics[width=1\linewidth]{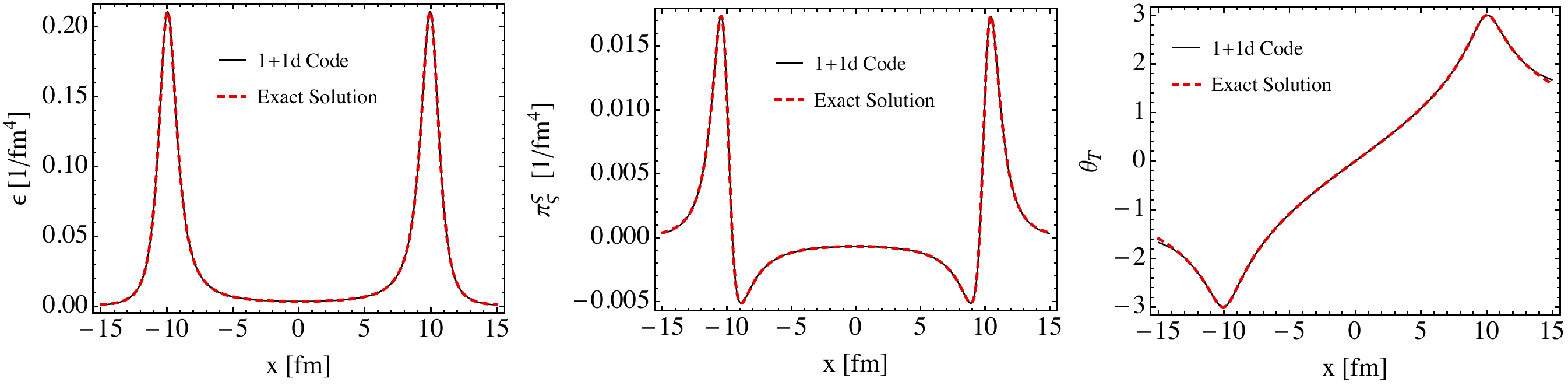}
\caption{Comparison of the numerical 1+1d solution to the second-order viscous hydrodynamics equations with the corresponding exact Gubser-flow solution \cite{Marrochio:2013wla}  at $\tau = 10$ fm/c.  The initial central temperature was $T_0 = 600$ MeV at $\tau_0 = 1$ fm/c with $4\pi\eta/s = 3$.  For this test we took $q = (1\;{\rm fm/c})^{-1}$.}
\label{fig:vhydro-comp}
\end{figure}

For the second-order hydrodynamic approximation subject to Gubser flow one has to solve two coupled ordinary differential equations subject to a boundary condition at $\rho{\,=\,}\rho_0$. For the case of Israel-Stewart viscous hydrodynamics, the necessary equations are~\cite{Marrochio:2013wla}
\begin{eqnarray}
&&\frac{1}{\hat{T}}\frac{d\hat{T}}{d\rho }+\frac{2}{3}\tanh \rho =
\frac{1}{3}\bar{\pi}_{\varsigma }^{\varsigma }(\rho )\,\tanh \rho \,, 
\label{eq:istemp}
\\
&&\frac{d\bar{\pi}_{\varsigma }^{\varsigma }}{d\rho }
+\frac{4}{3}\left( \bar{\pi}_{\varsigma }^{\varsigma }\right)^{2}\tanh \rho
+\frac{\bar{\pi}_\varsigma^\varsigma}{\hat{\tau}_\pi} =\frac{4}{15}\tanh\rho \,,
\label{eq:isshear}
\end{eqnarray}
where $\bar{\pi}_{\varsigma}^{\varsigma} \equiv \hat{\pi}_\varsigma^\varsigma/(\hat{T}\hat{s})$ and $\hat{\tau}_\pi = 5\eta/(s\hat{T})$.  Once the solution of these ordinary differential equations is obtained, one can map them back to Minkowski space using Eqs.~(\ref{eq:desitter1}) and (\ref{eq:desitter2}).

We present our results for the comparison of our Israel-Stewart solver and the exact solution in Fig.~\ref{fig:vhydro-comp}.  We use the same algorithm, lattice spacing, and temporal time step as in the main body of the text.  As one can see from Fig.~\ref{fig:vhydro-comp}, we are able to obtain excellent agreement between our numerical solution of the 1+1d Israel-Stewart partial differential equations and the exact solution subject to Gubser flow, even as late as $\tau = 10$ fm/c.  There are some small discrepancies near the boundary of the simulated region which are due to boundary effects.  We have checked that these effects can be reduced by using larger simulation volumes.

\subsection*{Gubser flow using anisotropic hydrodynamics}
\label{app:gubserahydro}

The dynamical equations needed to describe the de Sitter-space evolution of a system subject to Gubser flow using anisotropic hydrodynamics are~\cite{Nopoush:2014qba}
\ba
4\frac{d\log\hat\lambda}{d\rho}+\frac{3 \hat\alpha_\varsigma^2\left(\frac{{\cal H}_{2
   L}(\bar{y})}{{\cal H}_2(\bar{y})}+1\right)-4}{3\hat\alpha_\varsigma^2-1} \, \frac{d\log\hat\alpha_\varsigma}{d\rho}+ \tanh\rho\left(\frac{{\cal H}_{2T}(\bar{y})}{{\cal H}_2(\bar{y})}+2\right) &=& 0\, ,
\label{eq:1st-mom-final2} \\
\frac{6\hat\alpha_{\varsigma }}{1-3 \hat\alpha _\varsigma ^2} \frac{d \hat\alpha_\varsigma}{d\rho} -\frac{3 \left(3 \hat\alpha_\varsigma^4-4\hat\alpha_\varsigma^2+1\right)}{4\hat\tau_{\rm eq} \hat\alpha _{\varsigma }^5} \left(\frac{\hat{T}}{\hat\lambda}\right)^5+2\tanh\rho=0 \, ,
\label{eq:2nd-mom-final2}
\ea
where $\bar{y} \equiv \hat\alpha_\varsigma/\hat\alpha_\theta = \sqrt{(3\hat\alpha_\varsigma^2-1)/2}$. The ${\cal H}$-functions appearing above are defined in Eqs.~(\ref{eq:H2})-(\ref{eq:H2L}).  The set of equations can be closed by using the dynamical Landau matching condition
\ba
\hat{T}=\frac{\hat\alpha_\varsigma}{\bar{y}}\left(\frac{{\cal H}_2(\bar{y})}{2}\right)^{1/4} \hat\lambda .
\label{eq:matching-final2}
\ea
Once the solution of these ordinary differential equations is obtained, one can map them back to Minkowski space using Eqs.~(\ref{eq:desitter1}) and (\ref{eq:desitter2}).

\begin{figure}[t]
\includegraphics[width=1\linewidth]{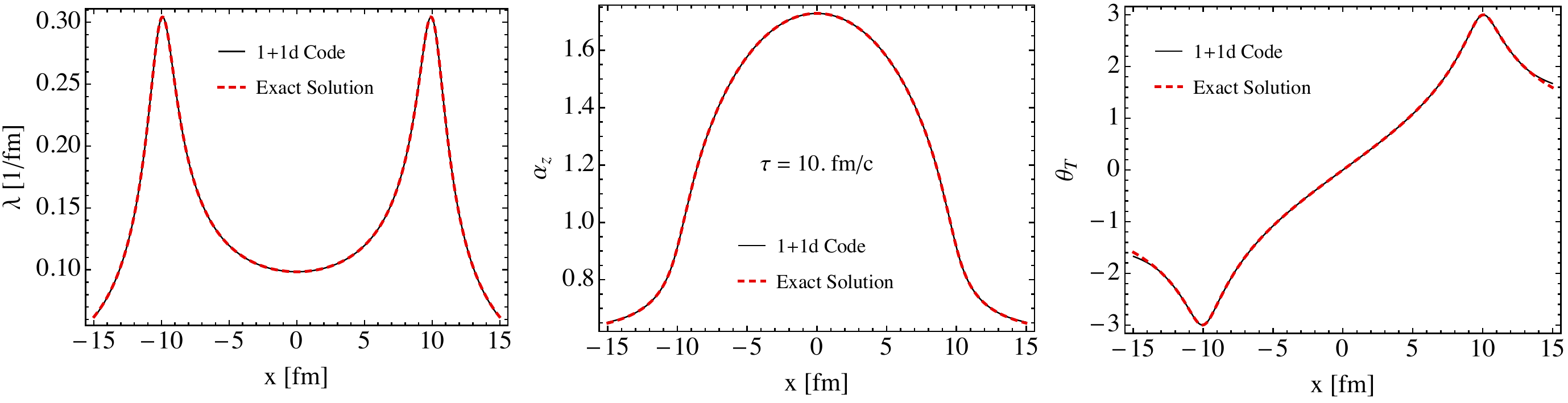}
\caption{Comparison of the numerical 1+1d solution to the anisotropic hydrodynamics equations with the corresponding exact Gubser-flow solution \cite{Nopoush:2014qba} at $\tau = 10$ fm/c.  The parameters, grid spacings, etc. used are the same as in Fig.~\ref{fig:vhydro-comp}.}
\label{fig:ahydro-comp}
\end{figure}

We present our results for the comparison of our anisotropic hydrodynamics solver and the exact solution in Fig.~\ref{fig:ahydro-comp}.  As before, we use the same algorithm, lattice spacing, and temporal time step as in the main body of the text.  As one can see from Fig.~\ref{fig:ahydro-comp}, we are able to obtain excellent agreement between our numerical solution of the 1+1d anisotropic hydrodynamics equations and the exact solution subject to Gubser flow, even as late as $\tau = 10$ fm/c.  As in the case of the Israel-Stewart solver, there are some small discrepancies near the boundary of the simulated region which are due to boundary effects.  We have, once again, checked that these effects can be reduced by using larger simulation volumes.

\bibliography{ahydro1p1}

\end{document}